\documentclass[reprint,aps, prd,letterpaper,showpacs,showkeys,twocolumn,floatfix,nofootinbib,superscriptaddress]{revtex4-1} 

\usepackage{amsmath,amssymb,amsfonts,graphicx}
\usepackage{bm}
\usepackage{slashed}
\usepackage{subfigure}
\usepackage[svgnames]{xcolor}

\newcommand{\vect}[1]{\boldsymbol{#1}}
\def\mathbi#1{\textbf{\em #1}}

\newcommand{\al}[1]{\begin{align} #1 \end{align}}
\newcommand{\non}{\nonumber}

\newcommand{\tV}{\text{V}}

\newcommand{\Ge}{\mathrm{GeV}}
\newcommand{\Gs}{\mathrm{GeV}^2}

\begin{document}

\title{Dihadron Fragmentation Functions within the NJL-jet Model}

\author{Hrayr~H.~Matevosyan}
\affiliation{CSSM and ARC Centre of Excellence for Particle Physics at the Tera-scale,\\ 
School of Chemistry and Physics, \\
University of Adelaide, Adelaide SA 5005, Australia
\\ http://www.physics.adelaide.edu.au/cssm
}

\author{Anthony~W.~Thomas}
\affiliation{CSSM and ARC Centre of Excellence for Particle Physics at the Tera-scale,\\ 
School of Chemistry and Physics, \\
University of Adelaide, Adelaide SA 5005, Australia
\\ http://www.physics.adelaide.edu.au/cssm
}

\author{Wolfgang Bentz}
\affiliation{Department of Physics, School of Science,\\  Tokai University, Hiratsuka-shi, Kanagawa 259-1292, Japan
\\ http://www.sp.u-tokai.ac.jp/
}

\begin{abstract}
Dihadron Fragmentation Functions (DFF) provide a vast amount of information on the intricate details of the parton hadronization process. Moreover, they provide a unique access to the "clean" extraction of nucleon transversity parton distribution functions in semi inclusive deep inelastic two hadron production process with a transversely polarised target. The NJL-jet model has been extended for calculations of light and strange quark unpolarised DFFs to pions, kaons and several vector mesons. This is accomplished by using the probabilistic interpretation of the DFFs, and employing the NJL-jet hadronization model in the Monte Carlo simulations that includes the transverse momentum of the produced hadrons. The strong decays of the vector mesons and the subsequent modification of the pseudoscalar meson DFFs are also considered. The resulting pseudoscalar meson DFFs are strongly influenced by the decays of the relevant vector mesons. This is because of the large combinatorial factors involved in counting the number of the hadron pairs that include the decay products. The evolution of the DFFs from the model scale to a typical experimental scale has also been performed.
\end{abstract}

\preprint{ADP-13-19/T839}
\pacs{13.60.Hb,~13.60.Le,~13.87.Fh,~12.39.Ki}
\keywords{Dihadron fragmentation functions, NJL-jet model, Monte Carlo simulations}

\date{\today}                                           

\maketitle

\section{Introduction}
\label{intro}

 Semi-inclusive deep-inelastic scattering (SIDIS) process, where a hard probe scatters off a target and is detected in the final state along with a single hadron, has served as the primary tool for exploring the three dimensional structure of hadrons. It has also permitted flavour separation of parton distribution functions (PDF) originally measured in fully inclusive deep inelastic scattering experiments (no observed final hadron). Moreover, SIDIS also allows one to extract the so-called transversity PDF through measurements of the single spin asymmetry in an experiment with a transversely polarised target. Such extractions rely on the factorisation theorems for the SIDIS cross-section, that express it as a convolution of a PDF, hard probe-parton scattering and  parton fragmentation functions (FF). Thus, a detailed knowledge of transverse momentum dependent (TMD) fragmentation functions is essential for the reliable extraction of transversity PDFs from SIDIS. Currently, such extractions rely on several approximations and assumptions, as our knowledge of both integrated and especially TMD FFs is still limited. This has been demonstrated in the recent experimental results from SIDIS measurements in HERMES~\cite{Airapetian:2012ki} and COMPASS~\cite{Makke:2013bya, Makke:2013dya}, where the existing parametrizations of the fragmentation and parton distribution functions failed to give an adequate description of the measured multiplicities for pions and  kaons. 
 
 In the extraction of TMDs, the behaviour of FFs in phenomenological parametrizations of experimental data is most often assumed  to be Gaussian, with a width that does not depend on the light-cone momentum fraction, the flavour of the fragmenting quark or the type of the produced hadrons. The recent results from HERMES~\cite{Airapetian:2012ki} and COMPASS~\cite{Adolph:2013stb} show that such simplistic approximations give an unacceptable description the data.

 An alternate method for extracting the transversity PDF was proposed in Refs.~\cite{Collins:1993kq,Jaffe:1997hf, Bianconi:1999cd, Bianconi:1999uc, Radici:2001na} utilising the SIDIS process with two final detected hadrons. It was realised that the corresponding cross section of such process can be factorized into the nucleon parton distribution function (PDF), a hard scattering cross-section and a dihadron fragmentation function (DFF) that describes the production of two hadrons in a parton hadronization process. Here, instead of the transverse momentum convolution of PDF and FF, which appears in single hadron SIDIS, when integrating over the transverse momenta of both hadrons we obtain a simple product of collinear PDF and a DFF that depends on the sum of the produced hadrons' light-cone momentum fractions and their invariant mass. Such a separation gives direct access to the nucleon transversity PDF through a measurement of a single spin asymmetry in SIDIS with transversely polarised target, though a knowledge of the corresponding DFFs is required. The first such extraction was performed in Ref.~\cite{Bacchetta:2011ip,Bacchetta:2012ty} using the SIDIS two hadron asymmetry measured by HERMES~\cite{Airapetian:2008sk}. This asymmetry is a ratio of a product of transversity PDF and an interference DFF (IFF) over a product of unpolarised PDF and DFF. Information on IFFs crucial for these extractions can be obtained by considering asymmetries in two back to back hadron pair production in $e^+e^-$ annihilation~\cite{Boer:2003ya}. Projections for such asymmetries were given in Ref.~\cite{Bacchetta:2008wb}, where SIDIS two hadron asymmetries were fitted using a spectator model, and making use of the evolution equations of the DFFs derived in Ref.~\cite{Ceccopieri:2007ip}. In Ref.~\cite{Courtoy:2012ry}, IFFs were extracted by fitting a parametrization form to recent $e^+e^-$  measurements by the BELLE collaboration~\cite{Vossen:2011fk}, using input from PYTHIA~\cite{Sjostrand:2006za} for the unpolarised DFFs that enter the relations for the asymmetries.  A perturbative calculation for DFFs at large invariant mass was performed in Ref.~\cite{Zhou:2011ba}. 
 
 The most recent non-perturbative model for DFFs was constructed in Ref.~\cite{Bacchetta:2006un} based on the spectator approach, where the model parameters were fixed by fitting the unpolarised DFFs for $\pi^+\pi^-$ pairs to Monte Carlo (MC) samples generated using PYTHIA~\cite{Sjostrand:2006za}. In these studies, due to the limited information on both unpolarised and interference DFFs, a number of assumptions and extrapolations were employed. Moreover, the extractions of unpolarised DFFs were conducted on Monte Carlo events generated by PYTHIA. They therefore depend on the particular models used for the parton hadronization and resonance decays (such as vector to pseudoscalar mesons). These decays, as shown in our earlier work~\cite{Matevosyan:2013nla} and will be again demonstrated here, have large effects on DFFs.
  
 DFFs describe the production of two hadrons in the parton hadronization process. They are more challenging in terms of  both theoretical description and experimental extraction than ordinary FFs. The theoretical models should give a detailed picture of the hadronization to final states in order to accurately describe  DFFs, as the leading hadron approximation typically used in FF models is insufficient here. It is easy to see that, even in the region of large total light-cone momentum fraction, choosing a pair where one leading hadron has a large light-cone momentum fraction of the fragmenting quark requires the second hadron in the pair to have small light-cone momentum fraction. Thus it can be produced at higher (subleading) order. Hence a model providing complete hadronization picture is needed for such studies, such as the Lund model~\cite{Sjostrand:1982fn} implemented in the PYTHIA event generator~\cite{Sjostrand:2006za,Sjostrand:2007gs}. The original studies of Field and Feynman, based on their  quark-jet model~\cite{Field:1976ve,Field:1977fa} of "collinear" DFFs that depend only on the light-cone momentum fractions of each hadron in the pair, have been recently extended within the NJL-jet model including the kaon production channel and also exploring their scale evolution~\cite{Casey:2012ux,Casey:2012hg}.
  
 Here we expand the NJL-jet model and the corresponding Monte Carlo (MC) software of~\cite{Ito:2009zc,Matevosyan:2010hh,Matevosyan:2011ey,PhysRevD.86.059904,Matevosyan:2011vj,Matevosyan:2012ga,Matevosyan:2012ms} to calculate unpolarised DFFs of light and strange quarks to several low-lying pseudoscalar and vector mesons. We accomplish this by using the probabilistic interpretation of DFFs and the complete quark hadronization description given by the NJL-jet model. We also study the strong two- and three-body decays of the vector mesons and their effects on the resulting pseudoscalar meson DFFs, where we reported the first results for $u\to\pi^+\pi^-$ in our earlier work Ref.~\cite{Matevosyan:2013nla}. We also study the scale evolution of the DFFs, crucial for comparing our low energy model results with those extracted from experiments in the deep-inelastic regime. 
  
 This paper is organised in the following way. In the next Section we briefly introduce the NJL-jet model and explain in detail our method of calculating the DFFs using MC methods. In Section~\ref{SEC_RES_NJL}  we  present our results for the NJL-jet model calculations of unpolarised DFFs. In Section~\ref{SEC_RES_EVOL}, we  briefly discuss the QCD evolution of DFFs and present the sample results of our model DFFs evolved to a typical experimental scale. This will be followed by Section~\ref{SEC_CONC} with conclusions and outlook.
 
\section{Calculating DFFs in the NJL-jet Model}
\label{SEC_CALC_DFF}

\subsection{The NJL-jet Model}
\label{SUB_SEC_NJL}

\begin{figure}[tb]
\centering
\includegraphics[width=0.9\columnwidth,clip]{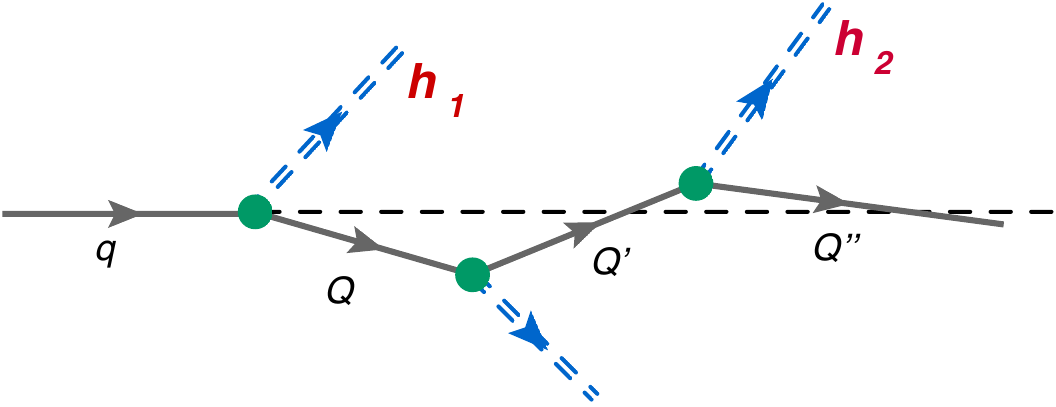}
\caption{The quark-jet hadronization mechanism.}
\label{PLOT_QUARK_JET}
\end{figure}
 
 The NJL-jet model provides a multi-hadron emission framework for describing the quark hadronization process, where any single hadron production probability is calculated within an effective quark model. The multi-hadron emission is described using the original quark-jet hadronization framework of Field and Feynman~\cite{Field:1976ve,Field:1977fa}, which is schematically depicted in Fig.~\ref{PLOT_QUARK_JET}. Here the fragmenting quark sequentially emits hadrons that do not re-interact with other produced hadrons or the remnant parton. The elementary hadron emission probabilities at each vertex are calculated using the effective quark model of Nambu and Jona-Lasinio (NJL)~\cite{Nambu:1961tp,Nambu:1961fr}, using the Lepage-Brodsky regularisation scheme with dipole cut-off, as described in Ref.~\cite{Matevosyan:2011vj}.  Using the interpretation of fragmentation functions as probability densities allows one to extract them from the corresponding hadron multiplicities. These are calculated as Monte Carlo averages of the hadronization process of a quark with a given flavour when restricting the total number of emitted hadrons for each fragmentation chain to a predefined number. In the limit of an infinite number of produced hadrons, used in the original formalism of Field and Feynman, this procedure yields a coupled set of integral equations for both FFs and DFFs~\cite{Field:1976ve,Field:1977fa}. A detailed study of convergence to this limit with an increasing number of produced hadrons has been performed in Ref.~\cite{Matevosyan:2011ey}, where it was shown that for collinear FFs only a few hadron emissions are needed saturate the limit. The momentum and quark flavour conservation is built-in to the NJL-jet model, and the resulting FFs satisfy all the relevant momentum and isospin sum rules~\cite{Matevosyan:2010hh,Matevosyan:2011ey}.
 
 The first study within the NJL-jet model of the collinear DFFs, that depend on the light-cone momentum fractions of each produced hadron, for light and strange quark to pion, kaon and mixed pairs was performed in Ref.~\cite{Casey:2012ux}. The QCD scale evolution of these DFFs was studied in Ref.~\cite{Casey:2012hg}. Here we tackle the extended DFFs that depend on the sum of the light-cone momentum fraction and the invariant mass of the produced hadron pair, which requires the knowledge of the transverse momenta of the produced hadrons. An extension of the NJL-jet model to describe the transverse momenta of the hadron produced in the fragmentation process was performed in Ref.~\cite{Matevosyan:2011vj}. That involved the calculation of the transverse momentum dependent elementary hadron emission probabilities, and tracking of the remnant quark transverse momentum in the hadronization process. Thus the same techniques are also used in this study.  
    
In this work we study the hadronization of both light and strange quarks to pseudoscalar $\pi$, $K$ and vector ($\rho$, $\omega$, $K^*$ and $\phi$) mesons.  The very important effects of the vector meson (VM) strong decays on DFFs of pseudoscalars has been demonstrated in Ref.~\cite{Matevosyan:2013nla}. Hence we pay attention to a detailed description of the most important two- and three- body strong decay channels of these vector mesons. Any weak or electromagnetic effect is excluded from our calculation, as fragmentation functions encode only the strong interaction physics. We use the previous results for VM elementary splitting functions calculated in Ref.~\cite{Matevosyan:2011ey,PhysRevD.86.059904}, where their two-body decays were also included.

\vspace{-0.7cm}
\subsection{Calculating DFFs using MC}
\label{SUB_SEC_MC}

 Here we describe the kinematics of the fragmentation of a quark of flavour $q$ to two hadrons $h_1$ and $h_2$, following the standard conventions of Ref.~\cite{Radici:2001na}. The coordinate system is chosen such that the $z$ axis is along the direction of the 3-momentum of the initial fragmenting quark $q$.  The momenta of the quark and the two produced hadrons $h_1$ and $h_2$ are denoted as\footnote{We use the following LC convention for Lorentz 4-vectors $(a^-,a^+,\mathbi{a}^\perp)$, $a^\pm=\frac{1}{\sqrt{2}}(a^0\pm a^3)$ and $\mathbi{a}^\perp = (a^1,a^2)$. }
\begin{align}      
\label{EQ_MOMENTA}
&P_q=(k^-,k^+,\vect{0}),\\ \nonumber
&P_{h1}\equiv P_1 = (z_1 k^-, P_1^+, \vect{P}_{1}^{\perp}),\ P_1^2 = M_{h1}^2, \\ \nonumber
&P_{h2}\equiv P_2 = (z_2 k^-, P_2^+, \vect{P}_{2}^{\perp}),\ P_2^2=M_{h2}^2,
\end{align}
where $z_1\equiv P_{h1}^-/k^-, M_{h1}$ and $z_2\equiv P_{h2}^-/k^-, M_{h2}$ are the corresponding light-cone momentum fractions and the masses of the hadrons. 

 Then we can express the invariant mass of the two hadrons in terms of their transverse momenta and momentum fractions:
\al{
\non
\label{EQ_INV_MASS}
&M_h^2=  (1+\kappa)M_{h1}^2  +\frac{(1+\kappa)}{\kappa}M_{h2}^2+  \frac{(\kappa \vect{P_{1}^{\perp}} - \vect{P_{2}^{\perp}})^2}{ \kappa} ,
\\ 
&\kappa\equiv {z_2}/{z_1}.
}

  There are kinematic endpoints for $z_1$ and $z_2$ for any given $M_h^2$. We can easily see from Eq.~(\ref{EQ_INV_MASS}), that to satisfy the non-negativity of the transverse momentum squared, we should have
\begin{align}
\label{EQ_Z_BOUND}
&M_h^2 - (1+\kappa)M_{h1}^2  - \frac{(1+\kappa)}{\kappa}M_{h2}^2 \geq 0.
\end{align}
 Then it is easy to see that $M_h^2$ will satisfy
\begin{align}
\label{EQ_Z_MIN_BOUND}
M_h^2 \geq (M_{h1} + M_{h2})^2,
\end{align}
with equality at
\al{
\kappa = \frac{M_{h2}}{M_{h1}}\ .
}

 Also, $z_1$ and $z_2$  cannot individually take values $0$ or $1$, as this would violate the momentum conservation or the on-shell condition for the produced hadrons. 
 
 We denote the unpolarised dihadron fragmentation functions as $D_q^{h_1h_2}(z,M_h^2)$. They are functions of the sum of the light-cone momentum fractions $z=z_1+z_2$ and the invariant mass squared $M_h^2=(P_1+P_2)^2$, of the produced hadron pair. We employ the number density interpretation for the $D_q^{h_1h_2}(z,M_h^2)$ to extract them by calculating the corresponding multiplicities using a MC average over simulations of the quark hadronization process, similar to the method employed for the single hadron FF extractions~\cite{Matevosyan:2011ey,PhysRevD.86.059904,Matevosyan:2011vj,Matevosyan:2012ga}:
\begin{align}
\label{EQ_MC_EXTRACT}
&D_q^{h_1h_2}(z,M_h^2)\ \Delta z\ \Delta M_h^2 \\ \nonumber
&= \left< N_{q}^{h_1 h_2}(z,z+\Delta z; M_h^2,M_h^2  +\Delta M_h^2 )\right>&\\ \nonumber
&\equiv  \frac{ \sum_{N_{Sims}}  N_{q}^{h_1 h_2}(z,z+\Delta z; M_h^2,M_h^2  +\Delta M_h^2 )} { N_{Sims} },
\end{align}
where $ \left< N_{q}^{h_1 h_2}(z,z+\Delta z; M_h^2,M_h^2  +\Delta M_h^2 )\right>$ is the average number of  hadron pairs, $h_1 h_2$, created with total momentum fraction in range $z$ to $z+\Delta z$ and invariant mass squared in range $M_h^2$ to $M_h^2  +\Delta M_h^2$. This average is calculated over $N_{Sims}$ Monte Carlo simulations of the hadronization process.  For each MC simulation we consider all the directly produced (so-called "primary") hadrons by the quark $q$ and calculate $z$ and $ M_h^2$ for all the hadron pairs, filling-in the corresponding histograms. We also repeat the procedure by now considering all the final state hadrons after allowing for the strong decays of the "primary" produced resonances, particularly the vector mesons. Throughout this paper we use $\Ge^{-2}$ units for unpolarised DFFs. We choose large enough $N_{Sims}$ and small enough $\Delta z$, $\Delta M_h^2$ to avoid significant numerical artefacts. 
  
\subsection{The Decay of Vector Mesons}
\label{SEC_VM_DECAY}

 An important aspect of modelling pseudoscalar meson DFFs is an accurate description of the produced resonance decays, as these strongly affect not only the magnitude but also the shape of these DFFs. In this work we include several vector meson production channels and we will only consider the predominant two- and three-body decays of these mesons. The relative branching ratios of different decay channels are taken to match those published by the Particle Data Group~\cite{Beringer:1900zz}.
 
 First we describe the two-body decays, that $\rho$, $\omega$, $K^*$ and $\phi$ mesons undergo. The differential decay rate of vector meson $V$ to $h_1 h_2$ pseudo-scalars can be expressed using a point-like coupling approximation as
\al{
\label{EQ_VM_2BODY_RATE}
d \Gamma = \frac{(2\pi)^4}{2 E_q}\left |\frac{g_{V}^{h_1 h_2} \epsilon^\mu (p_{2 \mu} -p_{1 \mu})}{D_V(q^2)} \right|^2 d \Phi^{(2)}(q,p_1,p_2),
}
where $g_{V}^{h_1 h_2}$ is the coupling constant, $D_V(q^2)$ is the inverse propagator of $V$ and $d\Phi^{(2)}(q,p_1,p_2)$ is the differential two-body phase space. The momenta of the vector and the final state mesons are  denoted by $q$, $p_1$ and $p_2$, the energy and the polarisation 4-vector of the decaying meson are $E_q $ and $\epsilon^\mu$.

 The momentum conservation and the on-mass-shell conditions for the above hadrons read: 
\al{
\label{EQ_DECAY_MOM_REL}
&q=p_1+p_2,\nonumber\\
&q^2=m_h^2,\quad
p_1^2=m_{h_1}^2,\quad 
p_1^2=m_{h_1}^2.
}
We also denote the fraction of the light-cone momentum of $V$ carried by $h_1$ as $z_1 \equiv p_{1}^-/q^- $ and the fraction carried by $h_2$ as $z_2 \equiv p_2^-/q^-$, where trivially $z_1+z_2=1$. 

 When summed over the polarisation of the decaying vector meson, the amplitude-squared in Eq.~(\ref{EQ_VM_2BODY_RATE}) depends only on the masses of the particles in the decay. Then the two-body phase space can be expressed as
\begin{align}
\label{EQ_PHASE_2BODY}
\non
d \Phi^{(2)}(z_1, \vect{p}_{1}^{\perp}) 
 =&\frac{d p_1^-  d^2 \vect{p}_1^\perp}{(2\pi)^3 2 p_1^-}\int \frac{d p_2^-  d^2 \vect{p}_2^\perp}{(2\pi)^3 2 p_2^-}  \delta^4(q-p_1-p_2)
\\
= &\frac{d z_1 d \varphi_1}{4(2\pi)^6}\  
\end{align}
where $\varphi_1$ is the azimuthal angle of $\vect{p}_1^\perp$ and the constraint
\al{\non
(\vec{p}_1^\perp -z_1\vec{q}^\perp)^2  + &z_1 z_2(q^\perp)^2
\\
& - z_1 z_2 m_h^2 + z_2 m_{h1}^2 + z_1 m_{h2}^2=0
 }
 is obtained from integration over the delta function. Thus the requirement that $(p_1^\perp)^2$ is positive yields the constraint on the possible range of values for $z_1$ and $z_2$
\al{
\label{EQ_2BODY_ENDPOINT}
z_1 z_2\ {m_h^2} - z_2 m_{h1}^2 - z_1 m_{h2}^2 \geq 0 
}

 To model the propagator of $\tV$, we use the Vector Meson Dominance framework and the parametrisation of invariant mass dependence of the vector meson resonance width used in extracting the properties of $\rho$ and $\omega$ mesons from the Spherical Neutral Detector experiment~\cite{Achasov:2001hb}.
\al{
D_\tV(m^2)&=m^2-m_\tV^2 + i m \Gamma_{\tV}(m),
}
where $m_V$ is the peak mass and $\Gamma_\tV(m)$ is the energy-dependent width of the VM. The absolute value squared of this form corresponds to the relativistic Breit-Wigner distribution.  General arguments using the angular momentum in the vicinity of the threshold impose a cubic dependence of  $\Gamma_\tV(m)$  on the three-momentum of the decay product pseudoscalar mesons in the rest frame of the VM. In experimental extractions the precise form of the energy dependence of $\Gamma_\tV(m)$ (that satisfy this constraint) is chosen so as to best describe the data. The parametrization of the width of the $\rho$ meson used in Ref.~\cite{Achasov:2001hb} was
\al{
\Gamma_\rho(m) &= \Gamma_\rho(m_\rho) \left[ \frac{q_\pi(m)}{q_\pi(m_\rho)} \right]^3 \left[ \frac{m_\rho}{m} \right]^2,
}
where the momentum of pions in the rest frame of the $\rho$ is
\al{
q_\pi &\equiv \frac{1}{2} \sqrt{m^2-4m_\pi^2}.
}
In the unequal mass case we find the centre of mass three-momentum is
\al{
q&=\frac{1}{2m}\sqrt{(m^2 - (m_1+m_2)^2) (m^2 - (m_1-m_2)^2)}.
}
The best fit values for the mass and the decay constant are
\al{
&m_\rho \approx 0.775~\mathrm{GeV},\\
&\Gamma_\rho(m_\rho) \approx 0.150\mathrm{GeV}.
}

 The three-body decay of the $\omega$ and $\phi$ mesons can be described using the isobar model, where the decay of the vector meson $\tV$ proceeds through intermediate $\rho$ meson production: $\tV \to \rho \pi \to 3 \pi$. We can use the parametrization of the amplitude for the process from Ref.~\cite{Achasov:2001hb}, keeping only the intermediate $\rho$ channel. Then, ignoring the small widths of the $\omega$ and $\phi$ mesons we can write the amplitude for the decay process as
\al{
\label{EQ_OMEGA}
M(p_1,p_2,p_3) =\varepsilon_{\mu\alpha\beta\gamma} \epsilon^\mu p_1^\alpha p_2^\beta p_3^\gamma  \sum_{i=0,\pm} \frac{g_{V \rho_i \pi}\ g_{\rho_i \pi \pi}}{D_{\rho_i}(v_i^2)} ,
}
where $\epsilon^\mu$ is the polarisation vector of the decaying meson, $p_{\{1,2,3\}}$ are the 4-momenta of the final pions (we can assign the indices according to the charge), $g_{V \rho_i \pi}$ is the vector meson~-~$\rho\pi$ coupling. $v_i$ is the invariant mass of the intermediate $\rho_i$ meson.

\begin{figure}[tb]
\centering 
\subfigure[] {
\includegraphics[width=0.9\columnwidth]{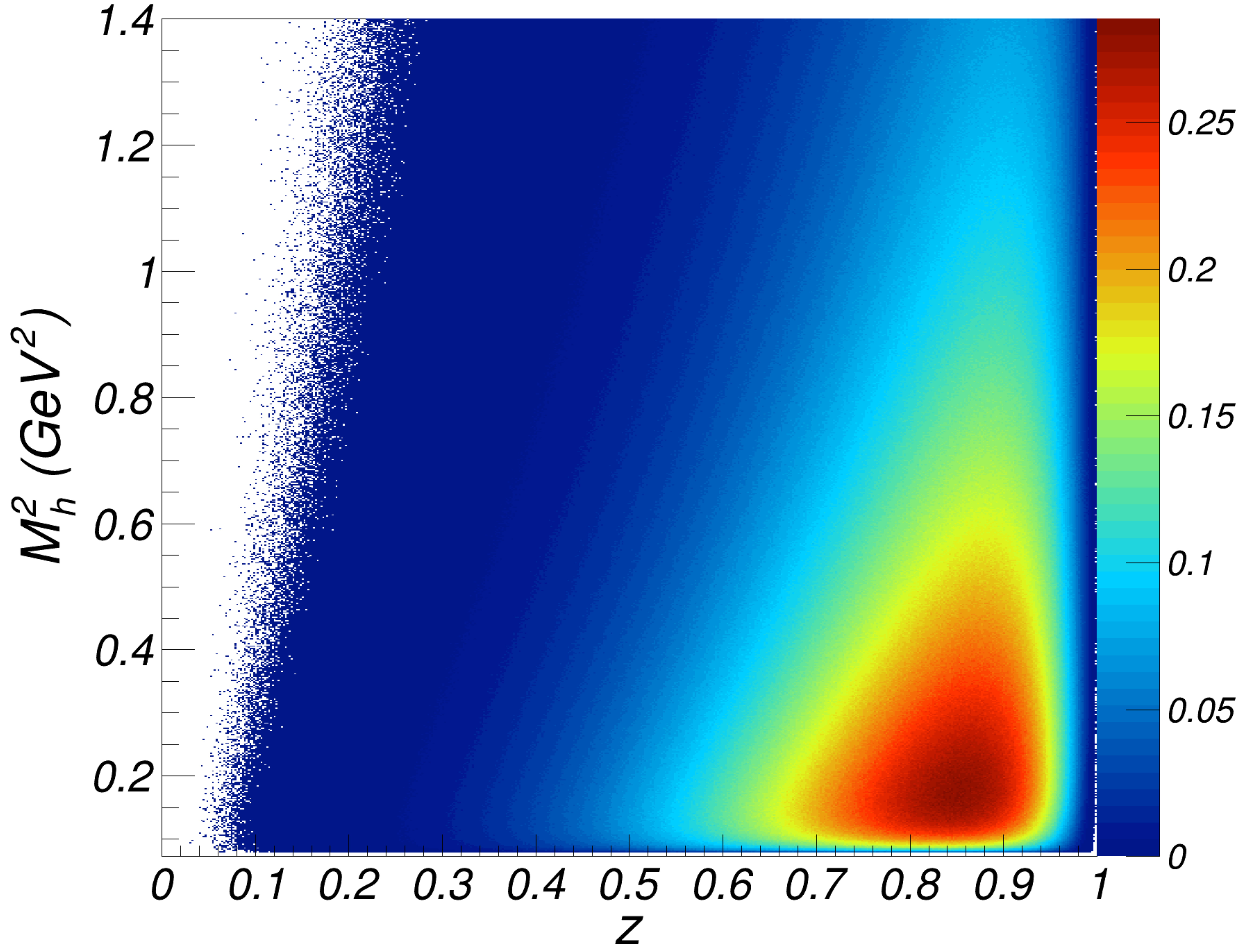}
}
\subfigure[] {
\includegraphics[width=0.9\columnwidth]{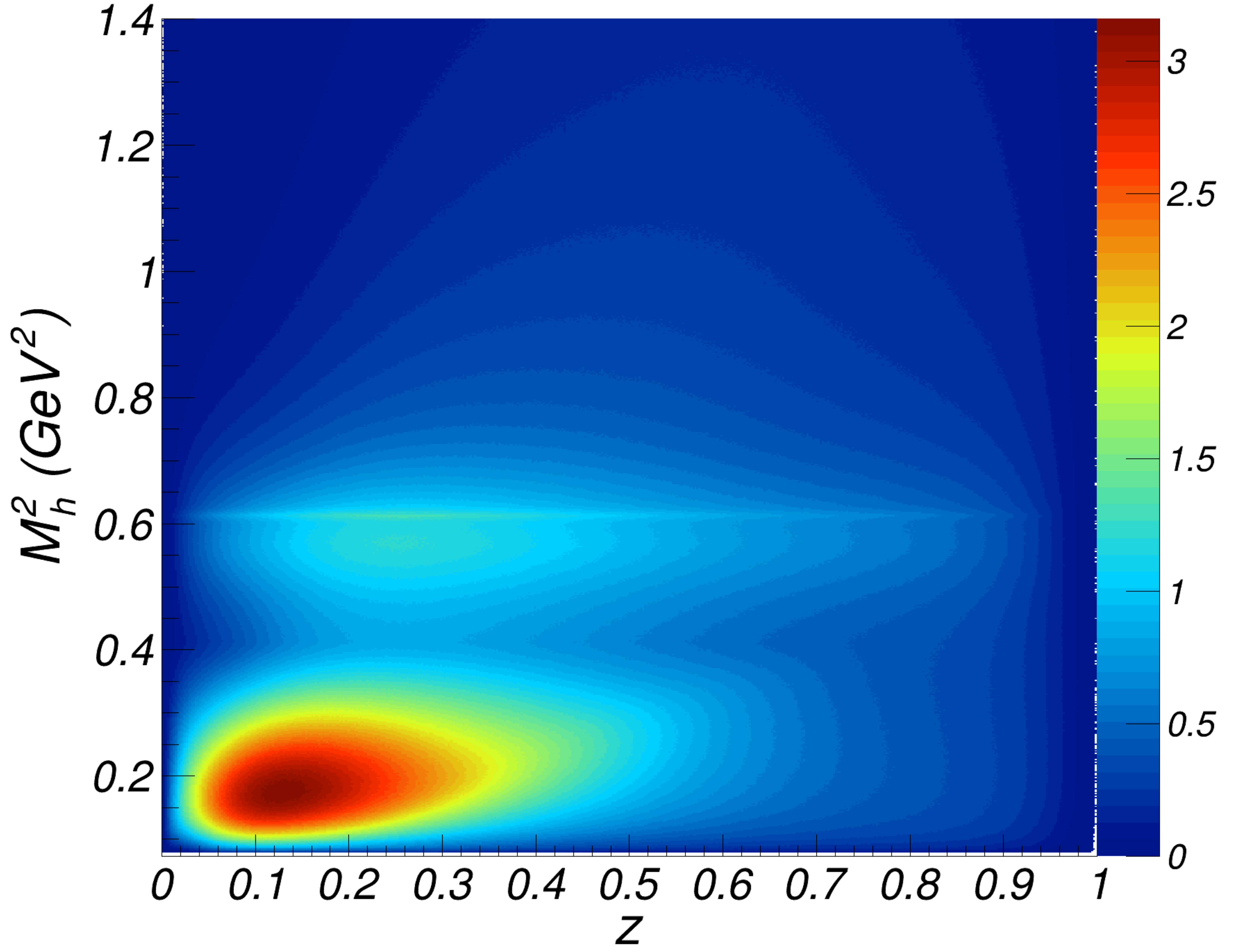}
}
\caption{The results for $D_u^{\pi^+\pi^-}$ for (a) only primary and (b) all the final state hadrons after decay of vector mesons from the NJL-jet MC simulations with $2$ primary produced hadrons.}
\label{PLOT_DFF_3D}
\end{figure}

The decay rate is then given by  
\al{
\label{EQ_VM_3PI_DECAY}
d \Gamma = \frac{(2\pi)^4}{2 E_q} |M(p_1,p_2,p_3)|^2 d \Phi^{(3)}&(q,p_1,p_2,p_3),
}
where $q$ is the 4-momentum and $E_q$ is the energy of the decaying meson. 

When summed over the polarisation of the decaying hadron, the absolute value square of the matrix element given in Eq.~(\ref{EQ_OMEGA}) depends only on the masses of the particles in the decay and the invariant masses of all possible pairs of the final state mesons. Then the  three-body phase space becomes independent of the Euler angles of the decay plane and can be expressed  as ~\cite{Beringer:1900zz}
\al{ \non
\int d \Phi^{(3)}&(q,p_1,p_2,p_3) 
\\
= &\frac{1}{2^4(2\pi)^7} \int_{(m_1+m_2)^2}^{(q-m_3)^2} d m^2_{12}\ \int_{m^2_{23,min}}^{m^2_{23,max}} d m^2_{23} 
}
where the subscript indices $ij$ denote the quantities related to the total four-momentum of  particles $i$ and $j$: $m_{12}^2 = (p_1+p_2)^2$. The expressions for $m^2_{23,min}$ and $m^2_{23,max}$ for a given value of $m^2_{12}$ can be found in Ref.~\cite{Beringer:1900zz}. Finally, we employ momentum conservation constraints and Monte Carlo techniques to sample the transverse momenta and the light cone momentum fractions of the produced final state pions according to the decay rate of \tV\  given in Eq.~(\ref{EQ_VM_3PI_DECAY}).
      
\begin{figure}[tb]
\centering 
\subfigure[] {
\includegraphics[width=0.9\columnwidth]{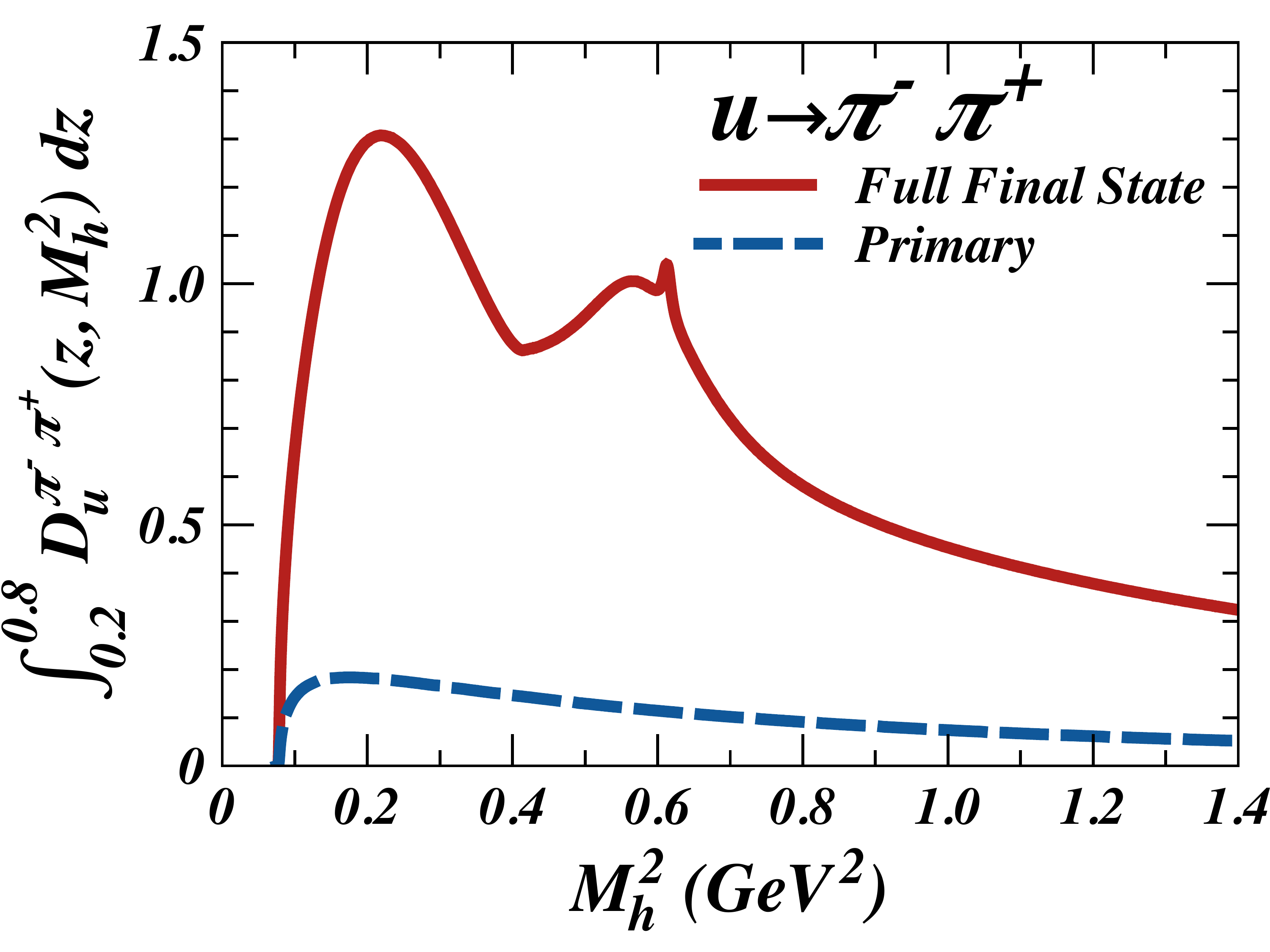}
}
\hspace{0.1cm} 
\subfigure[] {
\includegraphics[width=0.9\columnwidth]{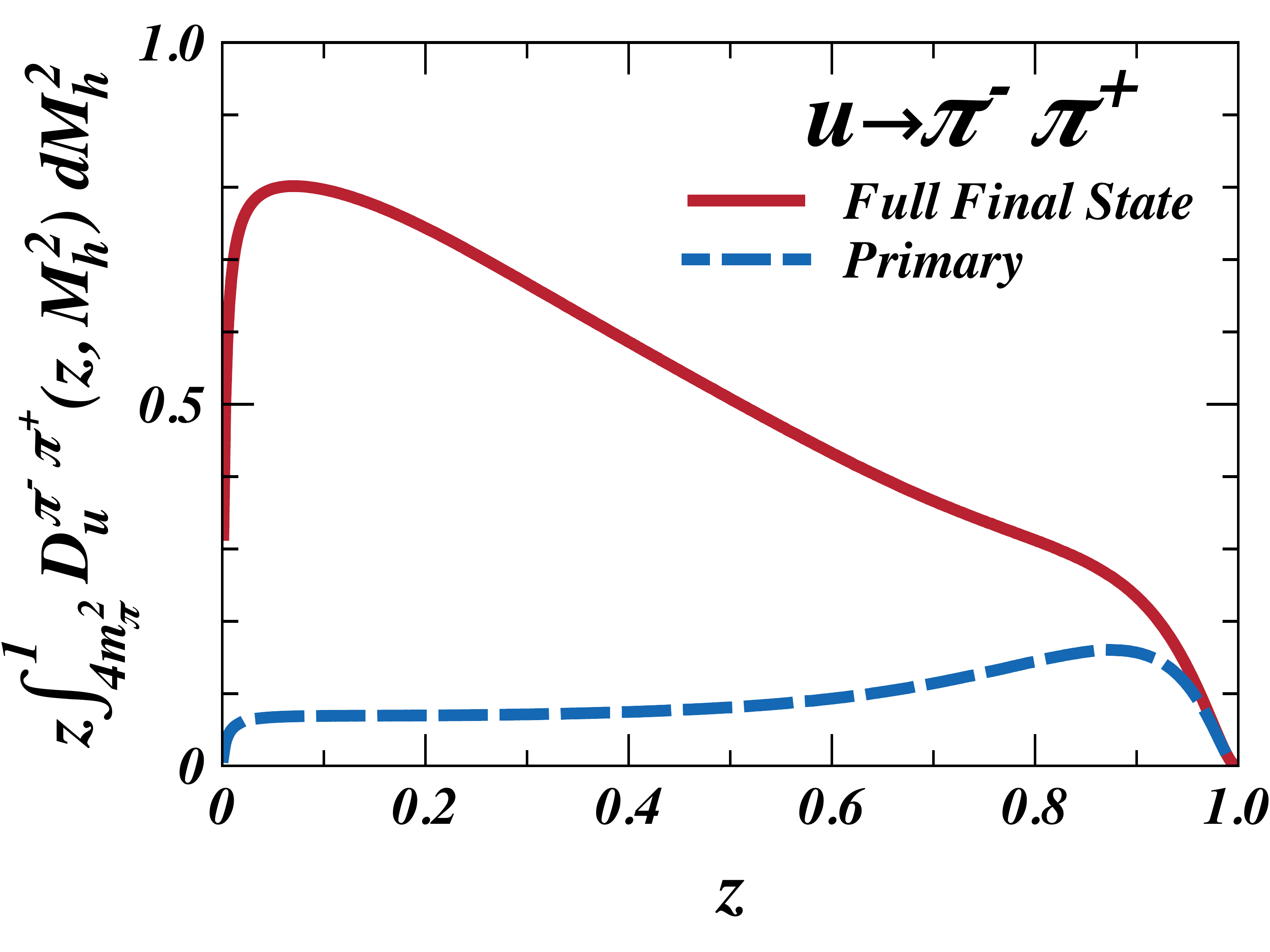}
}
\caption{The comparison of the results for $D_u^{\pi^+\pi^-}$ for the primary (blue dashed lines) and full (red solid lines) final state cases when (a) integrated over $z$ and (b) integrated over $M_h^2$. The MC simulations were performed with $8$ primary produced hadrons and over $10^{10}$ simulations.}
\label{PLOT_DFF_INT}
\end{figure}

\section{Results from the NJL-jet Model}
\label{SEC_RES_NJL}

\begin{figure}[tb]
\centering 
\subfigure[] {
\includegraphics[width=0.9\columnwidth]{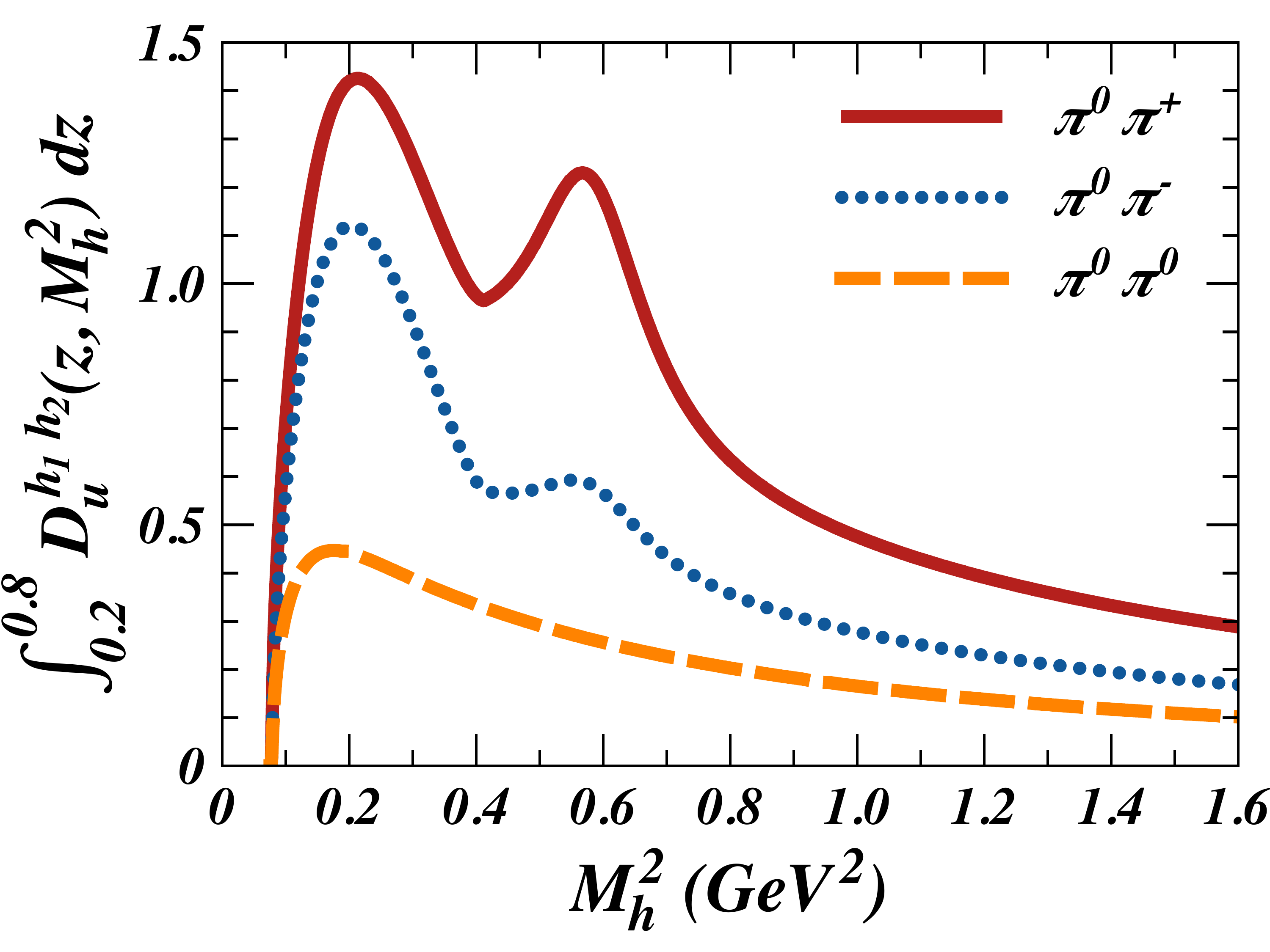}
}
\hspace{0.1cm} 
\subfigure[] {
\includegraphics[width=0.9\columnwidth]{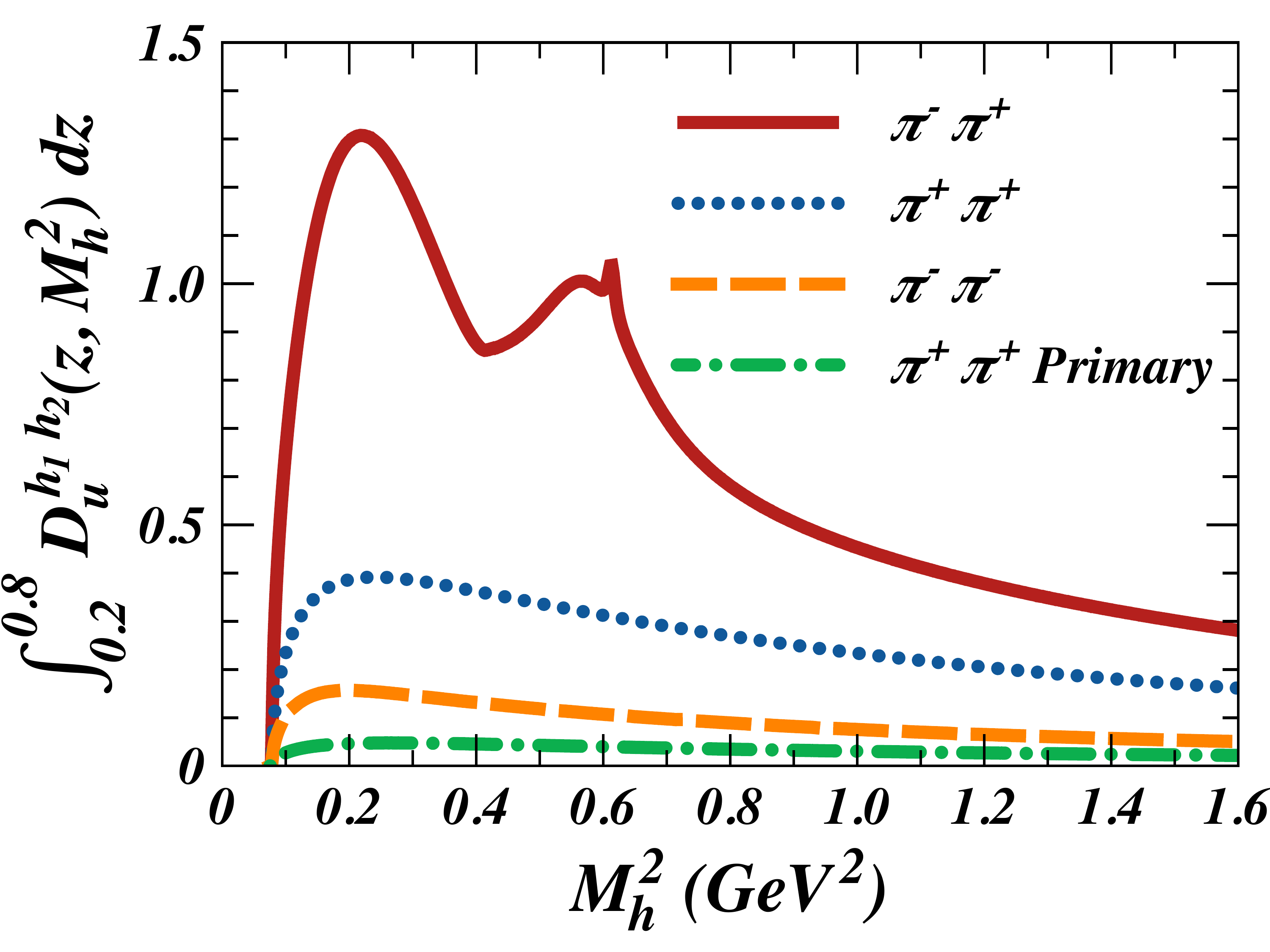}
}
\caption{The comparison of the results for $u$ quark DFFs to pion pairs with (a)  at least one neutral and (b) only charged pion pairs.}
\label{PLOT_DFF_U_PI_PI}
\end{figure}
\begin{figure}[tb]
\centering 
\subfigure[] {
\includegraphics[width=0.9\columnwidth]{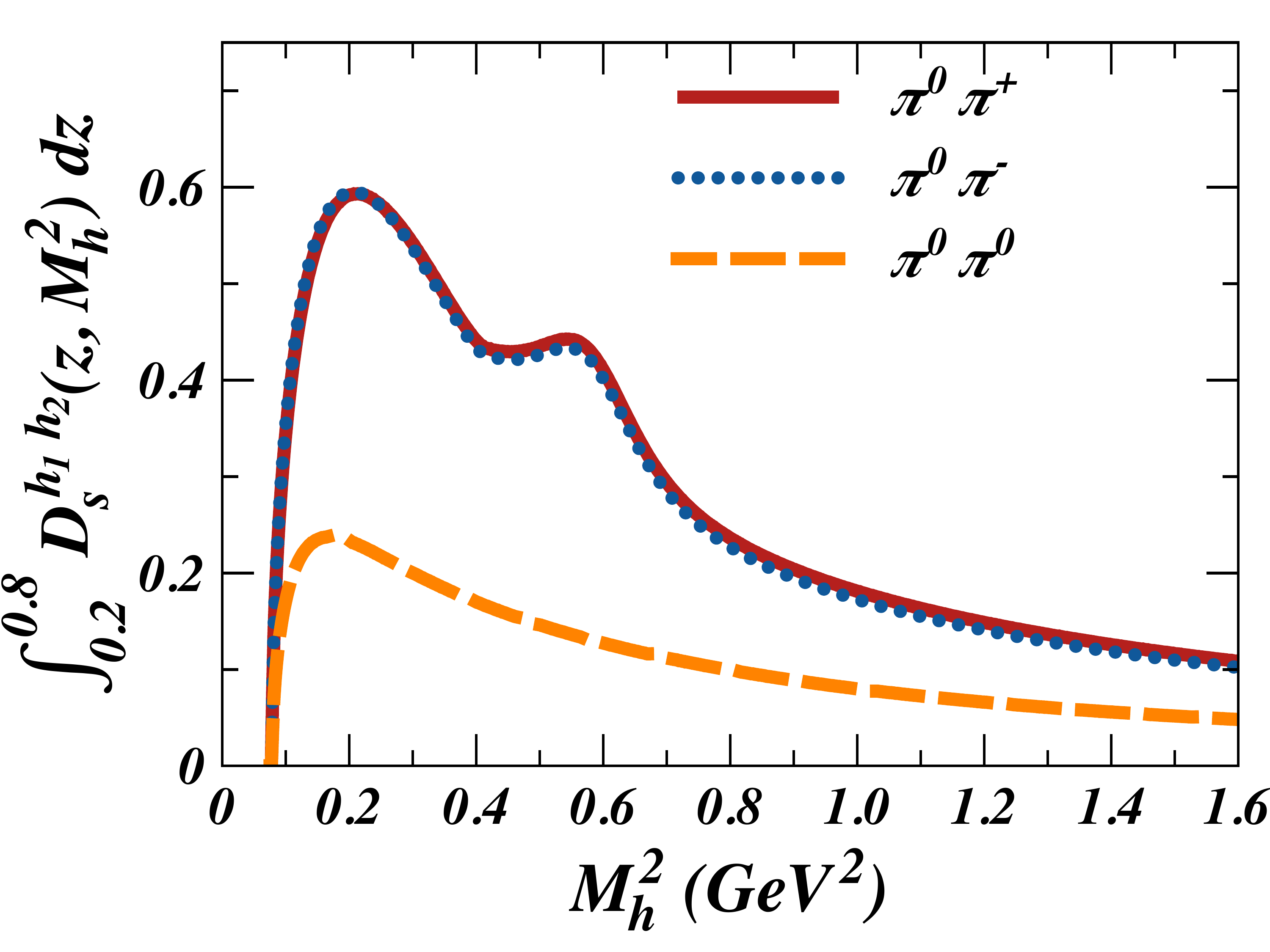}
}
\hspace{0.1cm} 
\subfigure[] {
\includegraphics[width=0.9\columnwidth]{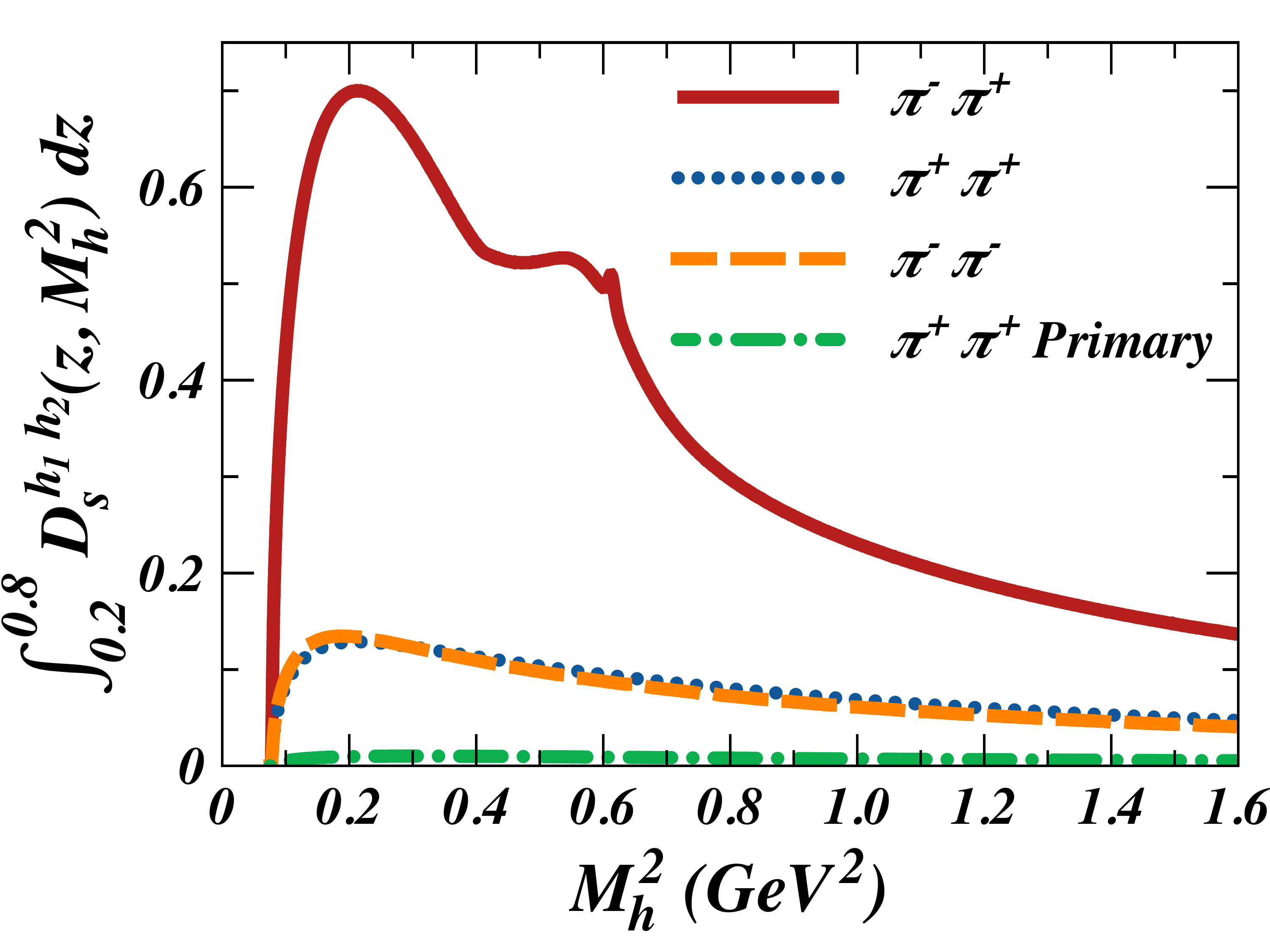}
}
\caption{The comparison of the results for $s$ quark DFFs to pion pairs with (a)  at least one neutral and (b) only charged pion pairs.}
\label{PLOT_DFF_S_PI_PI}
\end{figure}

 In this section we present the results of our calculations of DFFs of $u$, $d$ and $s$ quarks to pairs of mesons. For this, we performed MC simulations using our NJL-jet model based software framework and calculated the relevant DFFs using the formula in Eq.~(\ref{EQ_MC_EXTRACT}). The number of quark hadronization simulations $N_{Syms}=10^{10}$  and discretization sizes, $\Delta z=0.002$ and $\Delta M_h^2 = 2~\mathrm{MeV}^2$, were sufficient to avoid any sizeable numerical artefacts in the extracted functions. All the remaining model parameters are the same as in our previous work Ref.~\cite{Matevosyan:2011vj}
 
  The plots of the $D_u^{\pi^+\pi^-}(z,M_h^2)$ extracted from simulations with $2$ produced primary hadrons are shown in Fig.~\ref{PLOT_DFF_3D}, where the plot in Fig.~\ref{PLOT_DFF_3D}(a) depicts the results obtained when one only counts the pions produced directly by the fragmenting $u$ quark (primary hadrons), while Fig.~\ref{PLOT_DFF_3D}(b) depicts the results when the pions produced in the decays of the primary vector mesons (secondary hadrons) are also included. The distinct signature of the $\rho^0$ meson peak around $M_h^2\approx 0.775^2 \Gs $, the narrow peak of the $\omega\to \pi^+ \pi^-$ around $M_h^2\approx 0.783^2 \Gs $ (due to the small $\omega$-$\rho$ mixing effect), as well as the enhancement in the small $z$ and invariant mass region arising from $\omega$ meson decay are clearly visible. Our detailed studies, when we excluded the $\phi\to3\pi$ decay, showed that is has only a marginal effect on the DFFs of $u$ to pions, but contributes significantly to the DFFs of $s$ quark to pions. The dramatic effect on the shape and the magnitude of the DFF of including the full final state (FFS) was unexpected from the naive analogy to the moderate contribution of the secondary hadrons  to FFs (see Refs.~\cite{Matevosyan:2011ey,PhysRevD.86.059904}). It has been shown in our earlier publication~\cite{Matevosyan:2013nla}, that this behaviour is a general feature of DFFs by also analysing PYTHIA generated MC events,  and it can be explained using combinatorial factors. 

\begin{figure}[tb]
\centering 
\subfigure[] {
\includegraphics[width=0.9\columnwidth]{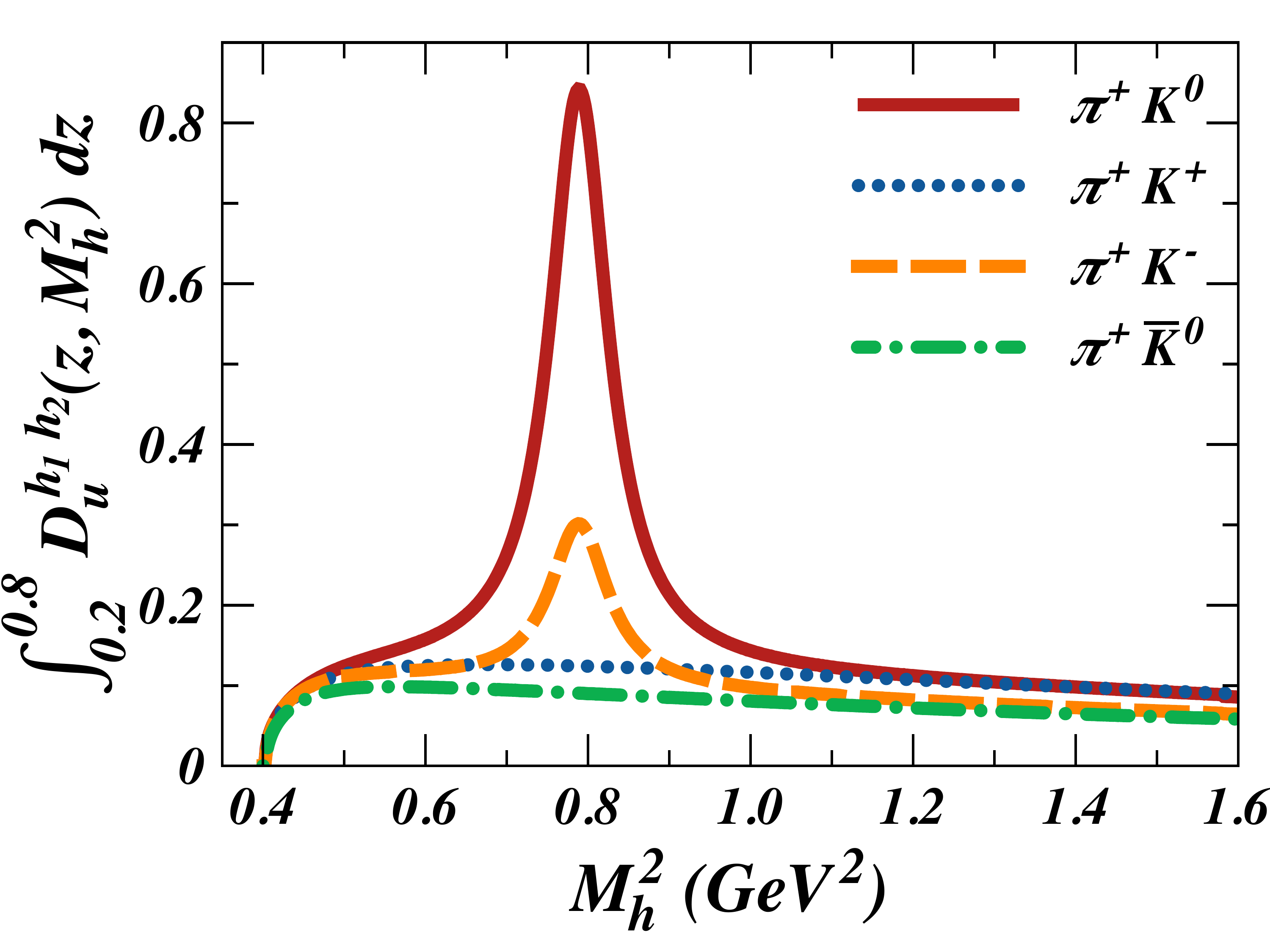}
}
\hspace{0.1cm} 
\subfigure[] {
\includegraphics[width=0.9\columnwidth]{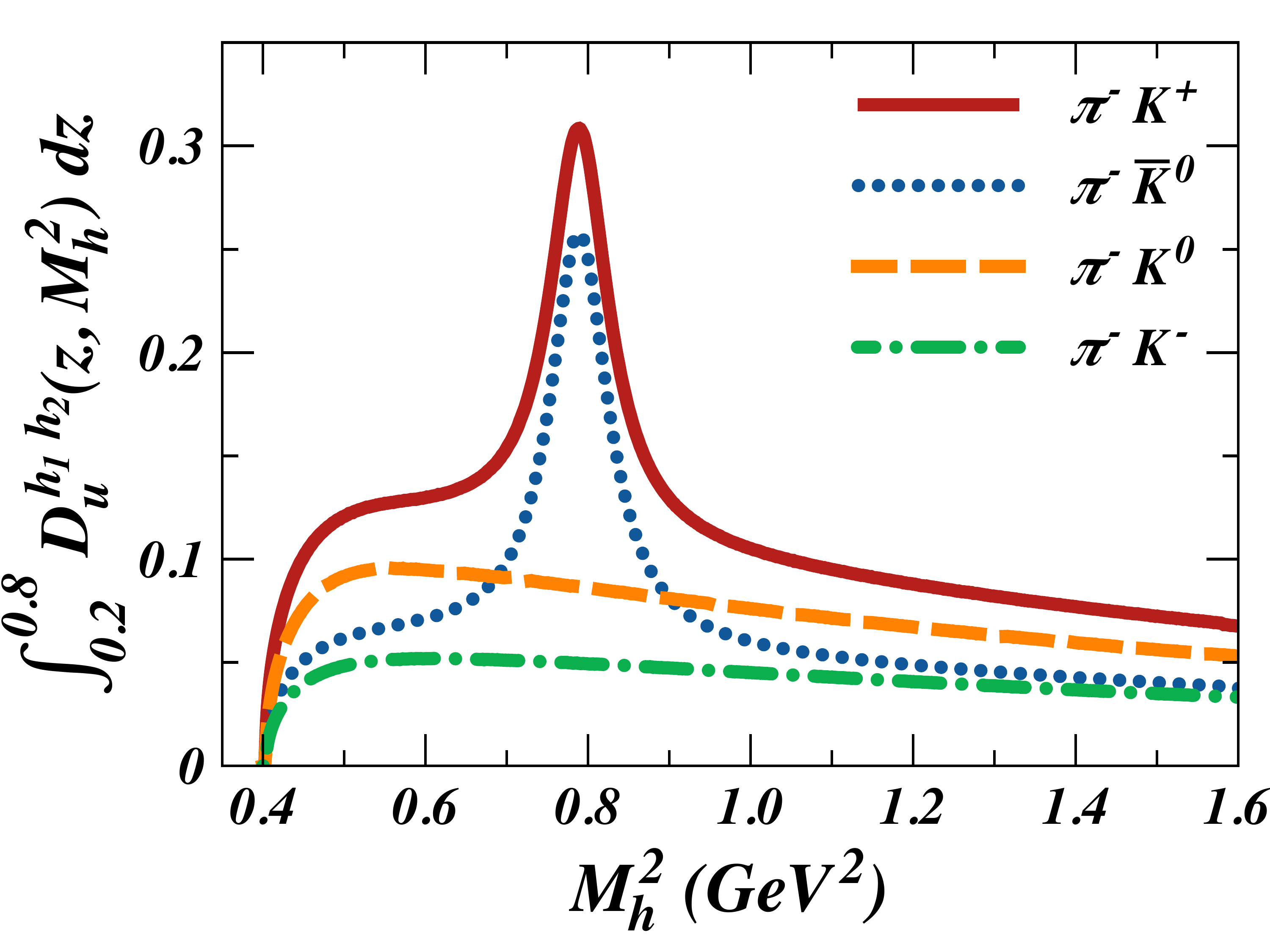}
}
\caption{The comparison of the results for $u$ quark DFFs to pairs with one kaon and (a) $\pi^+$  and (b) $\pi^-$,  integrated over $z$.}
\label{PLOT_DFF_U_PI_K}
\end{figure}

\begin{figure}[tb]
\centering 
\subfigure[] {
\includegraphics[width=0.9\columnwidth]{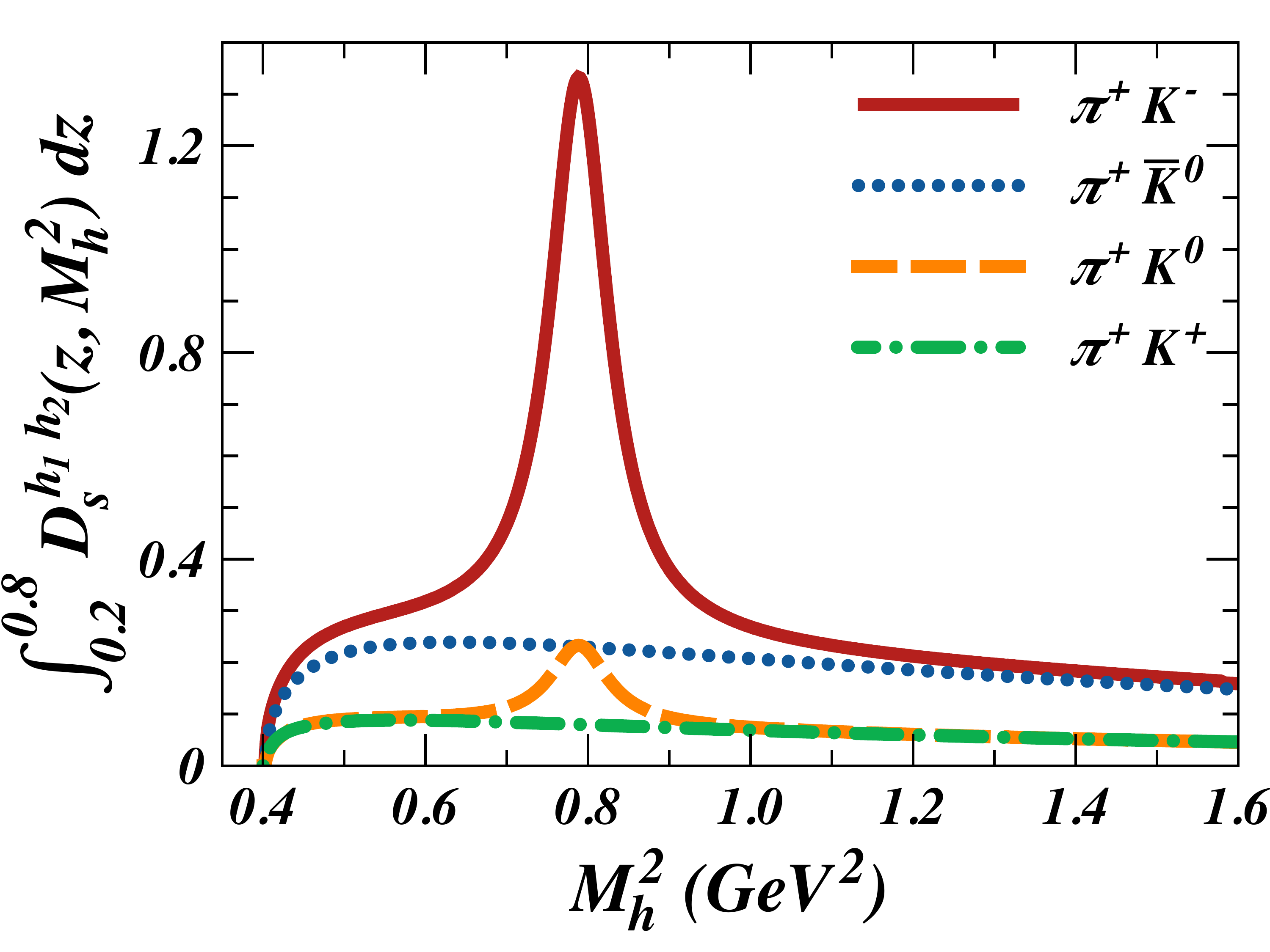}
}
\hspace{0.1cm} 
\subfigure[] {
\includegraphics[width=0.9\columnwidth]{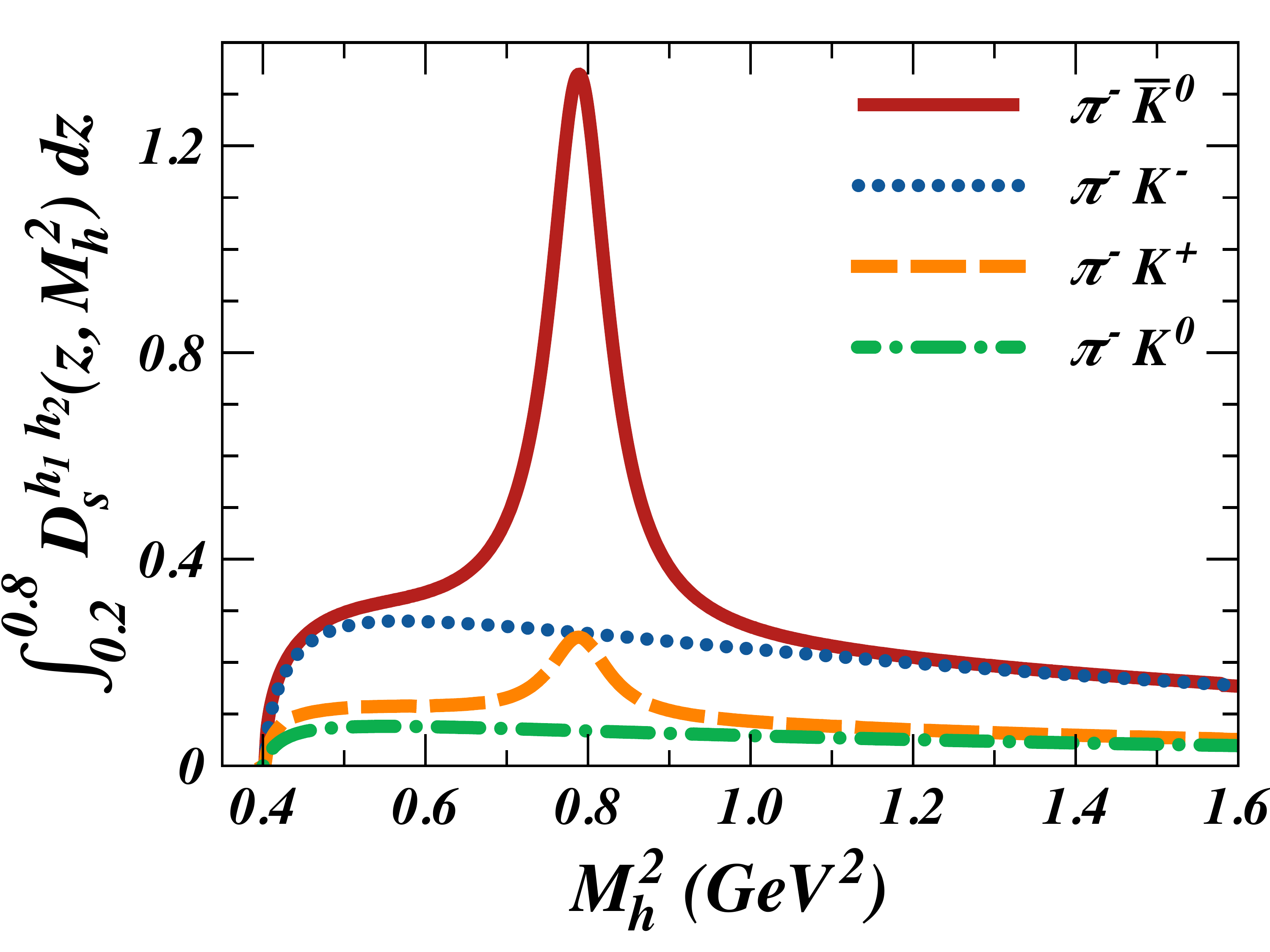}
}
\caption{The comparison of the results for $s$ quark DFFs to pairs with one kaon and (a) $\pi^+$  and (b) $\pi^-$,  integrated over $z$.}
\label{PLOT_DFF_S_PI_K}
\end{figure}

To investigate these features in detail  it is useful to integrate the DFF over some range of either $z$ or $M_h^2$. The plots in Fig.~\ref{PLOT_DFF_INT} depict the comparison of the results with primary and full set of final hadrons for $D_u^{\pi^+\pi^-}$, integrated over (a) $z$ in the region $0.2$ to $0.8$ to facilitate comparison with model calculations by Bacchetta et. al.~\cite{Bacchetta:2006un} and (b) over $M_h^2$ in the region $4m_\pi^2\approx 0.08~\Gs$ to $1~\Gs$. We readily see on the plots in Fig.~\ref{PLOT_DFF_INT}(a), depicting the invariant mass squared dependence of the DFF, the pronounced  effect of the vector meson resonances, such as the $\rho^0$ peak and the enhancement in the region below $0.4~\Gs$ coming from the $\omega\to \pi^+\pi^-\pi^0$ decay with shifted invariant mass due to unobserved $\pi^0$. 

 The plots in Figs.~\ref{PLOT_DFF_U_PI_PI} and~\ref{PLOT_DFF_S_PI_PI} depict the results for various pion pairs produced by a $u$ and an $s$ quark respectively. The peaks of the $\rho^+$ in the $\pi^0\pi^+$and the $\rho^0$ in the $\pi^-\pi^+$ channels are clearly evident, as well as the low invariant mass peak of the $\omega$ meson for $u$ quark and both $\omega$  and $\phi$ mesons for $s$ quark in channels with differently charged pions. The comparison of the primary and full final state DFFs for $\pi^+\pi^+$ pairs in Figs.~\ref{PLOT_DFF_U_PI_PI}(b) and~\ref{PLOT_DFF_S_PI_PI}(b) shows the strong enhancement effect due to the VM decays even in the channels not directly produced by these decays.

 The plots in Fig.~\ref{PLOT_DFF_U_PI_K} and~\ref{PLOT_DFF_S_PI_K} depict the results for the mixed pions and kaon pairs for both $u$ and $s$ quark fragmentations respectively. The pronounced $K^*$ peak is present in the $\pi^+K^-$ channel for both $u$ and $s$ quark DFFs. The relevant magnitudes of the direct contributions to DFFs can be intuitively deduced using favoured fragmentation channel arguments and the NJL-jet hadronization mechanism. For example, the suppression of the $s\to\pi^+K^+$ channel with respect to $s\to\pi^-K^-$, both of which have no direct contributions from resonance decays, can be understood using the argument that for an initial $s$ quark to produce the $\pi^-K^-$ pair in the final state, only one additional $\pi^+$ needs to be produced in the jet ($s\to u + K^- \to d + \pi^+ + K^- \to u +\pi^- +  \pi^+ + K^-$). For the $\pi^+K^+$ pair at least an emission of a $K^-$ by the initial $s$ quark (this splitting function is peaked at large $z$, as discussed in Ref.~\cite{Matevosyan:2012ga}) and a subsequent $\pi^-$ is needed ($s\to u + K^- \to d + \pi^+ + K^- \to u +\pi^- +  \pi^+ + K^- \to s +K^+ + \pi^- +  \pi^+ + K^-$). Thus the probability for the $\pi^-K^-$ to have at least half of the light-cone momentum of the initial quark is much less than for $\pi^+K^+$. 
 
\begin{figure}[tb]
\centering 
\subfigure[] {
\includegraphics[width=0.9\columnwidth]{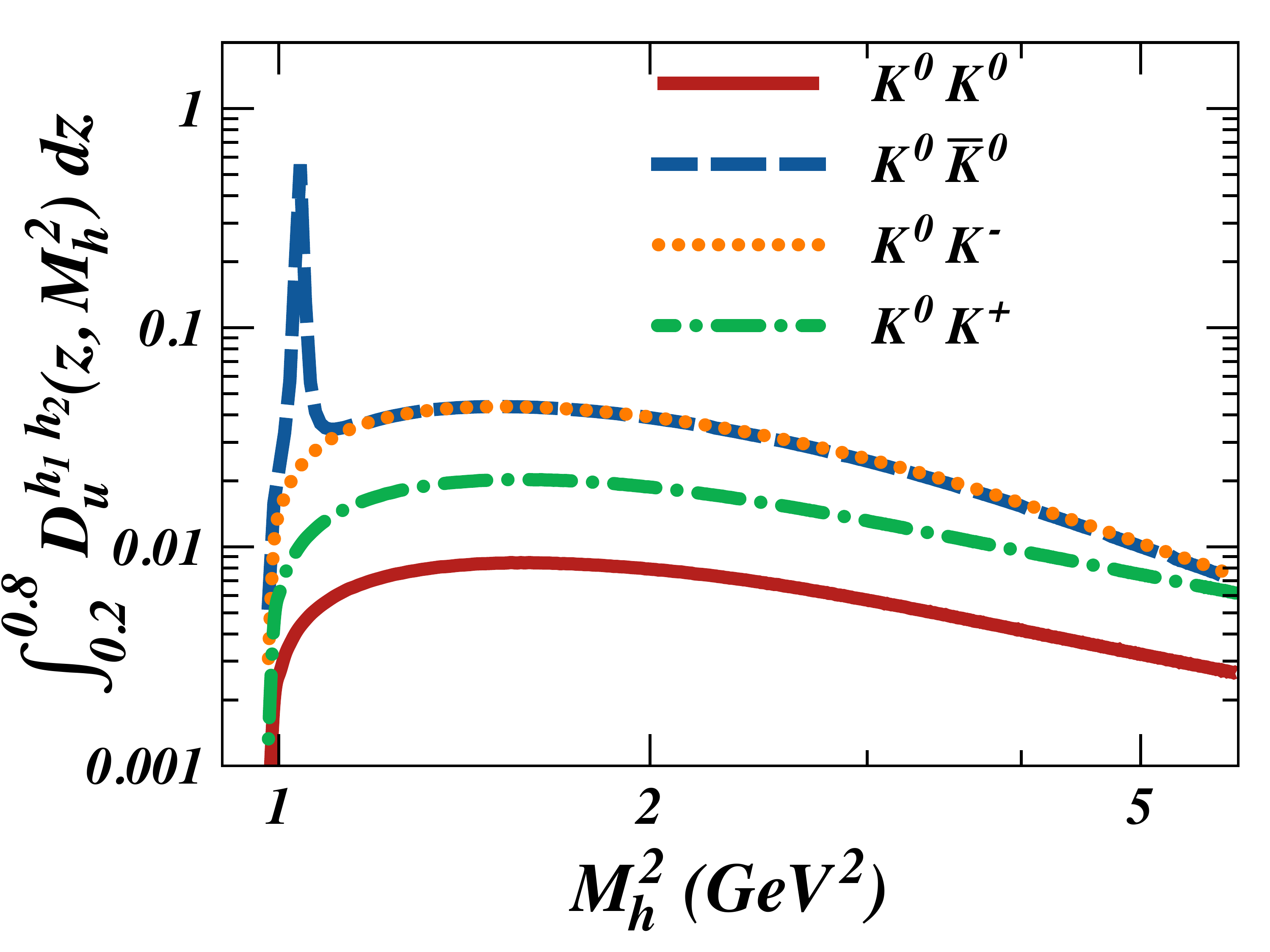}
}
\hspace{0.1cm} 
\subfigure[] {
\includegraphics[width=0.9\columnwidth]{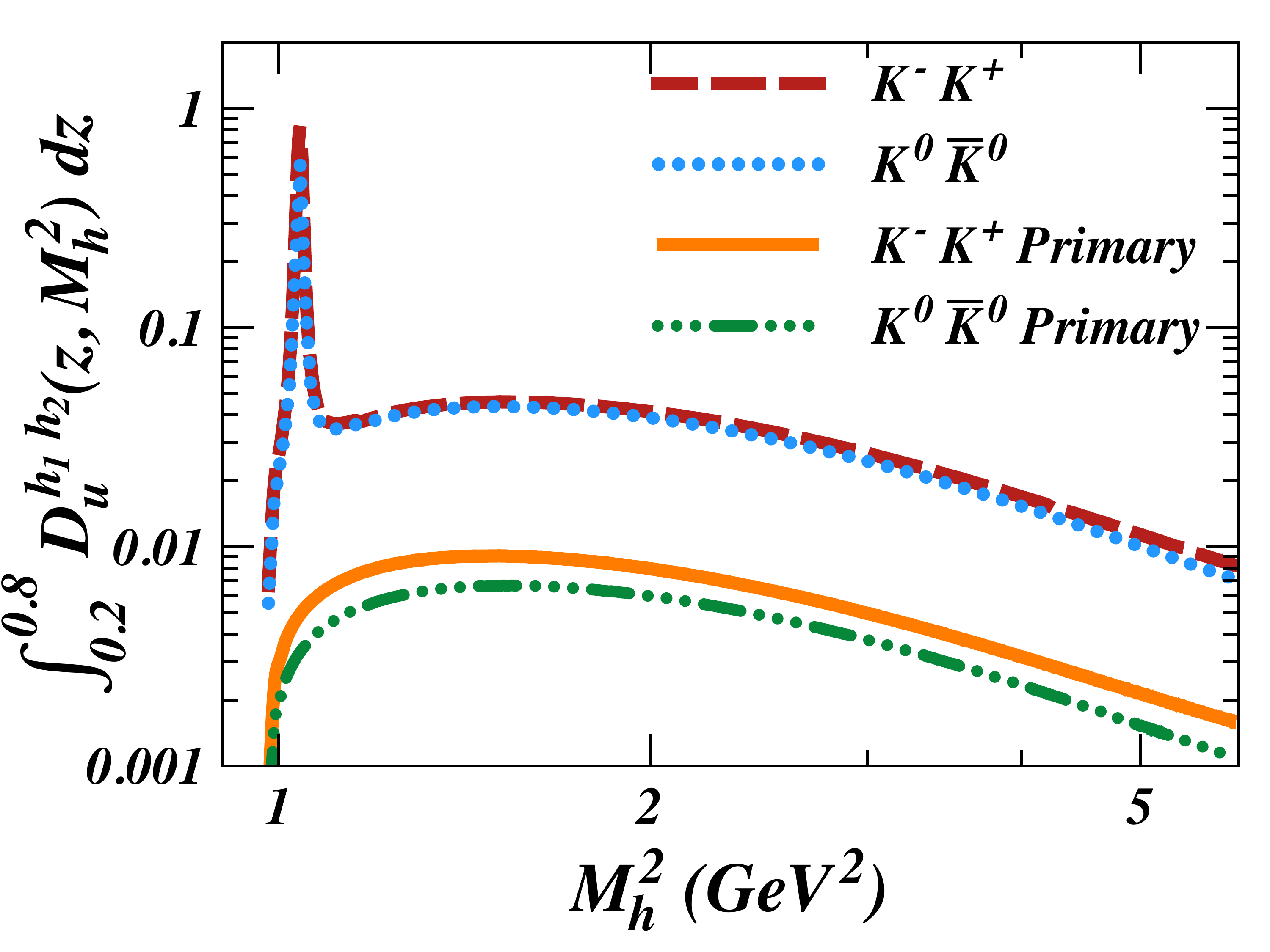}
}
\caption{The comparison of the results for DFFs to pion and kaon pairs for (a) $u$  and (b) $s$ quark, integrated over $z$. }
\label{PLOT_DFF_U_K_K}
\end{figure}
 
\begin{figure}[tb]
\centering 
\subfigure[] {
\includegraphics[width=0.9\columnwidth]{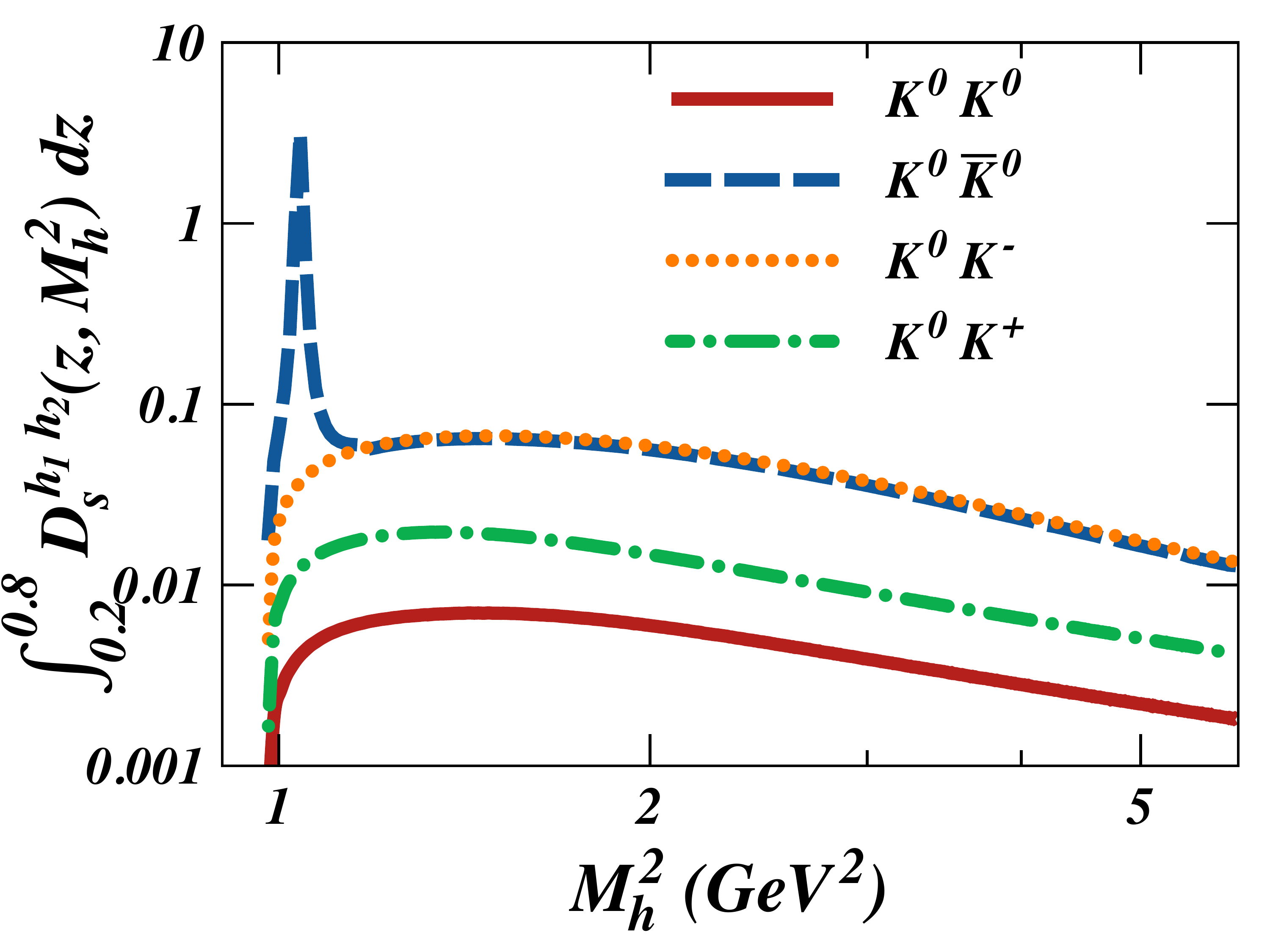}
}
\hspace{0.1cm} 
\subfigure[] {
\includegraphics[width=0.9\columnwidth]{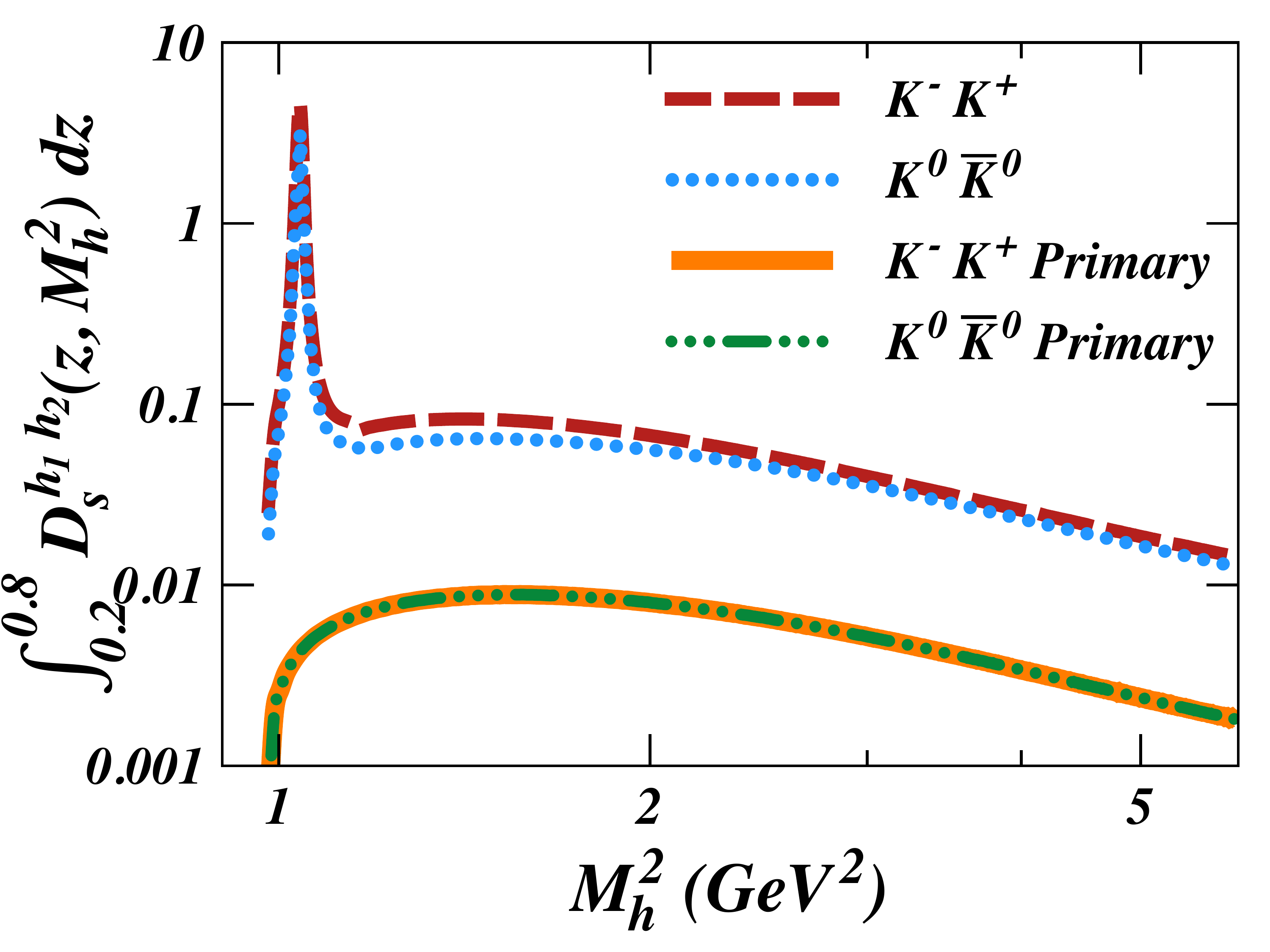}
}
\caption{The comparison of the results for DFFs to pion and kaon pairs for (a) $u$  and (b) $s$ quark, integrated over $z$.}
\label{PLOT_DFF_S_K_K}
\end{figure}

 The plots in Figs.~\ref{PLOT_DFF_U_K_K},~\ref{PLOT_DFF_S_K_K} depict the results for kaon pairs for both $u$ and $s$ quark fragmentations respectively. The $\phi$ meson peak is present in both $K^0 \bar{K}^0$ and $K^-K^+$ channels, where the plots in Figs.~\ref{PLOT_DFF_U_K_K}(b) and \ref{PLOT_DFF_S_K_K}(b) show their relative strength of $1:2$ as expected from isospin symmetry. Also, these plots show the large enhancement of DFFs from VM decays even far from the resonance peak region, by comparing the full final state DFFs with those produced only by primary mesons. The plots in Figs.~\ref{PLOT_DFF_U_K_K}(a) and \ref{PLOT_DFF_S_K_K}(a) show  that away from the resonance decay region,  $K^0 \bar{K}^0$ and $K^0K^-$ channels behave very similarly, as expected from a simplistic analysis of NJL-jet model using favoured hadron emission channel arguments, where the vector meson decay products simply enhance the DFF magnitudes without affecting the shapes.

 The vast amount of DFFs that include at least one resonance (before its strong decay of course) are not shown here for sake of brevity, but can be provided to the reader by the authors if requested.
 
\begin{figure}[tb]
\centering 
\includegraphics[width=0.9\columnwidth]{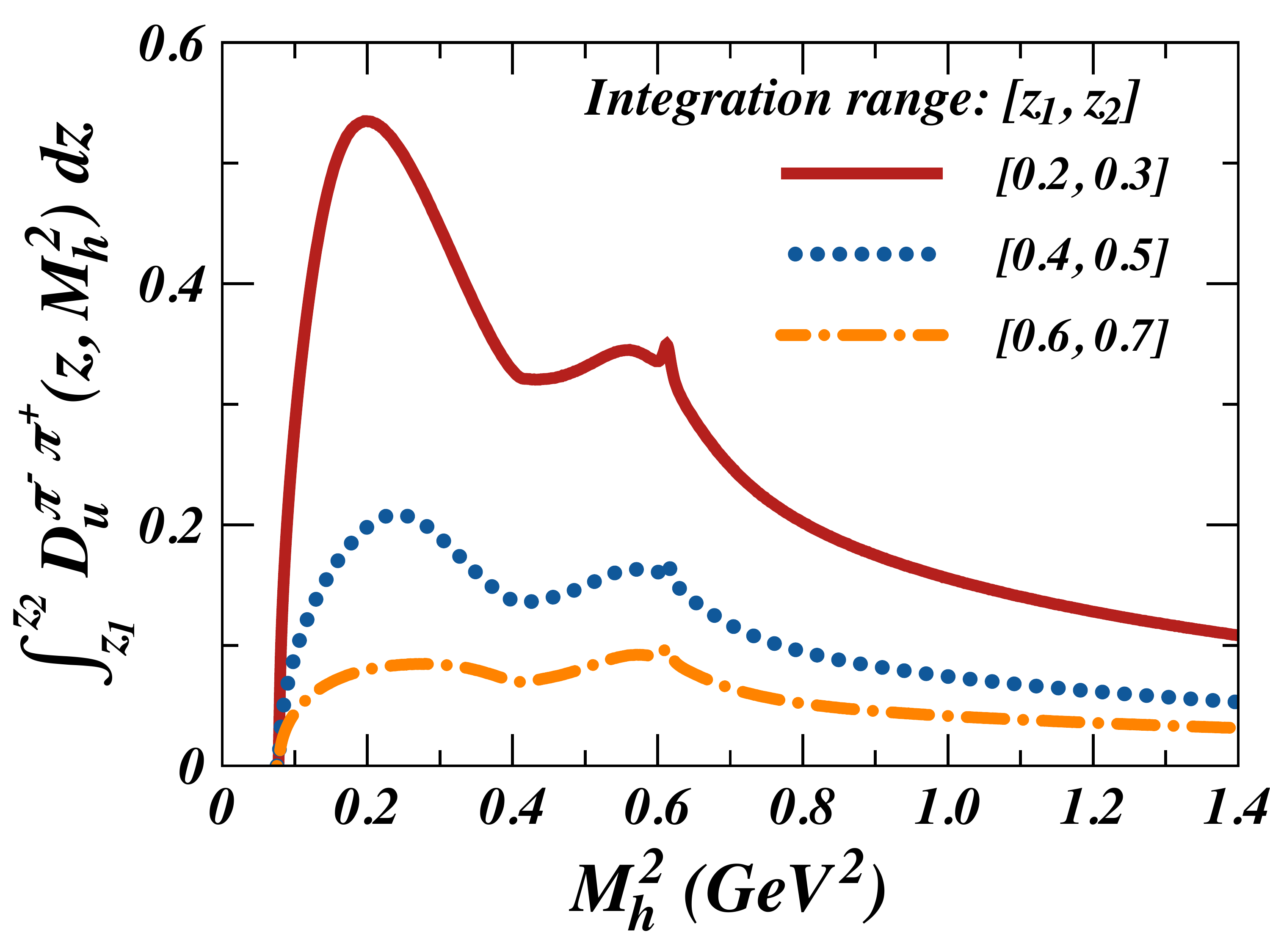}
\caption{$D_u^{\pi^+\pi^-}$ integrated over different ranges in $z$, as noted in the caption of the plot, versus $M_h^2$.}
\label{PLOT_DFF_U_PiMi_PiPl_SLICES}
\end{figure}

 To study the $z$ dependence of different resonance peak contributions to DFFs, we depict $D_u^{\pi^+\pi^-}$ integrated over several narrow ranges in $z$, shown in Fig.~\ref{PLOT_DFF_U_PiMi_PiPl_SLICES}. It is clear, that the $\omega$ meson contributions are predominant in the small $z$ region, while the $\rho$ meson peak gives the major contribution in the large $z$ region. This result can be easily deduced from the fact that even though $\rho$ and $\omega$ have the same elementary splitting functions in our model (modulo isospin factors), the $\pi^0$ from the three body decay of the $\omega$ will carry some of the light-cone momentum fraction of the parent, thus shifting the $\pi^-\pi^+$ pair to a lower $z$ region. Moreover, the low $z$ region is additionally enhanced by the numerous small-$z$ hadrons produced further in the quark hadronization process. 

\section{DFF Evolution}
\label{SEC_RES_EVOL}

 In this section we consider the QCD evolution of DFFs, which is paramount in order to compare our low energy model scale calculations (fixed to be $0.2~\Gs$ in the NJL-jet model) to the results obtained from experiment that have a typical energy scale of at least several $\Gs$. The leading order (LO) evolution equation of DFFs has been derived in Ref.~\cite{Ceccopieri:2007ip, Bacchetta:2008wb}
 \al{
 \label{EQ_DFF_EVOL}
 &\frac{d}{d\ \log Q^2}  D_{q}^{h_1,h2}(z,M_h^2,Q^2) 
 \\ \non
 & = \frac{\alpha_s(Q^2)}{2\pi}\sum_i\int_z^1 \frac{d u}{u}\ P_{qi}(u)\ D_{i}^{h_1,h2}\left(\frac{z}{u}, M_h^2, Q^2\right),
 }
where $\alpha_s(Q^2)$ is the strong coupling constant, the sum over $i$ denotes all the active quark flavours and gluon, and $P_{qi}(u)$ are the usual splitting functions of a quark of flavour $q$ to a parton $i$ carrying its light-cone momentum fraction $u$. 
 
\begin{figure}[t]
\centering 
\subfigure[] {
\includegraphics[width=0.9\columnwidth]{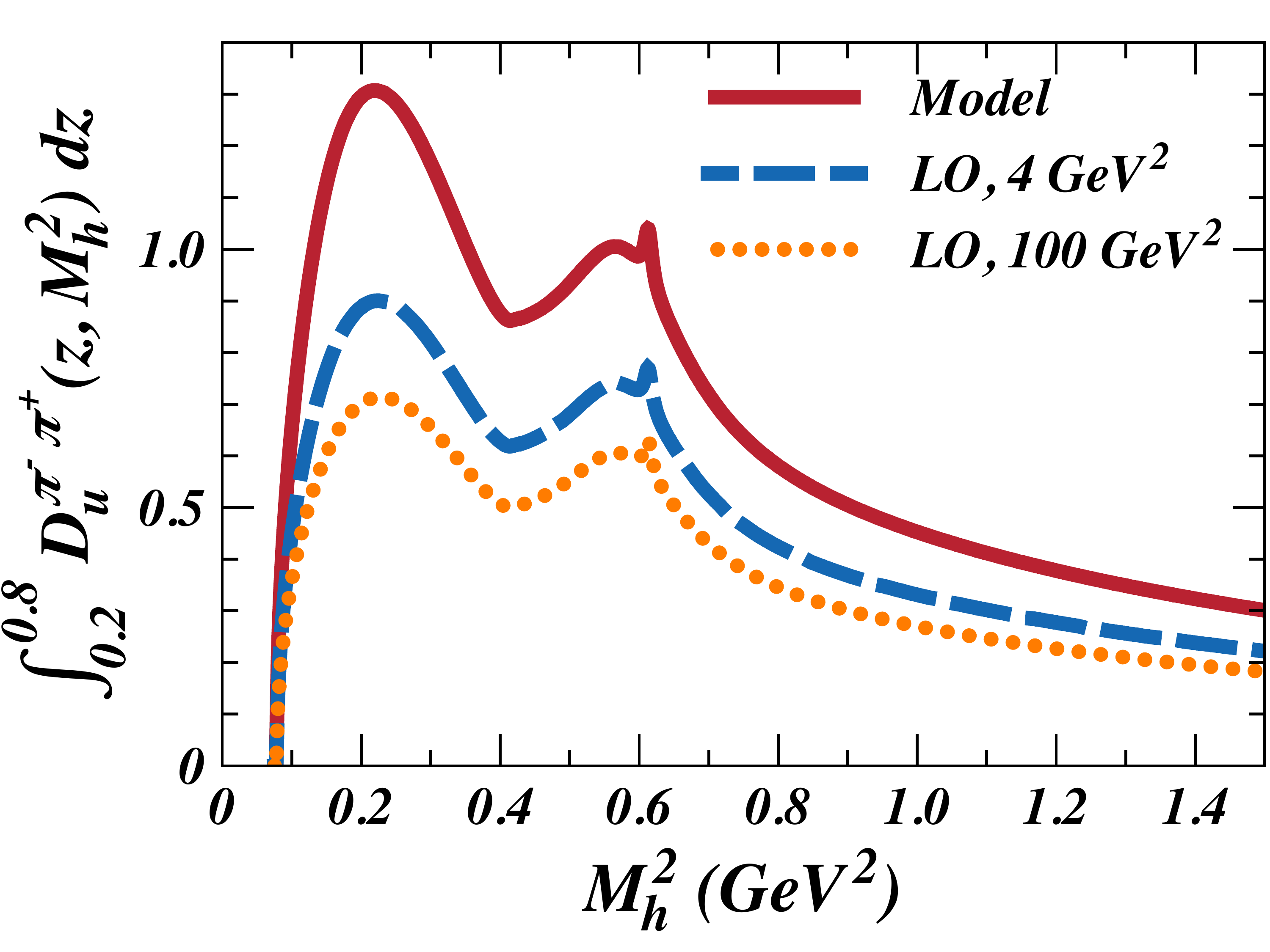}
}
\hspace{0.1cm} 
\subfigure[] {
\includegraphics[width=0.9\columnwidth]{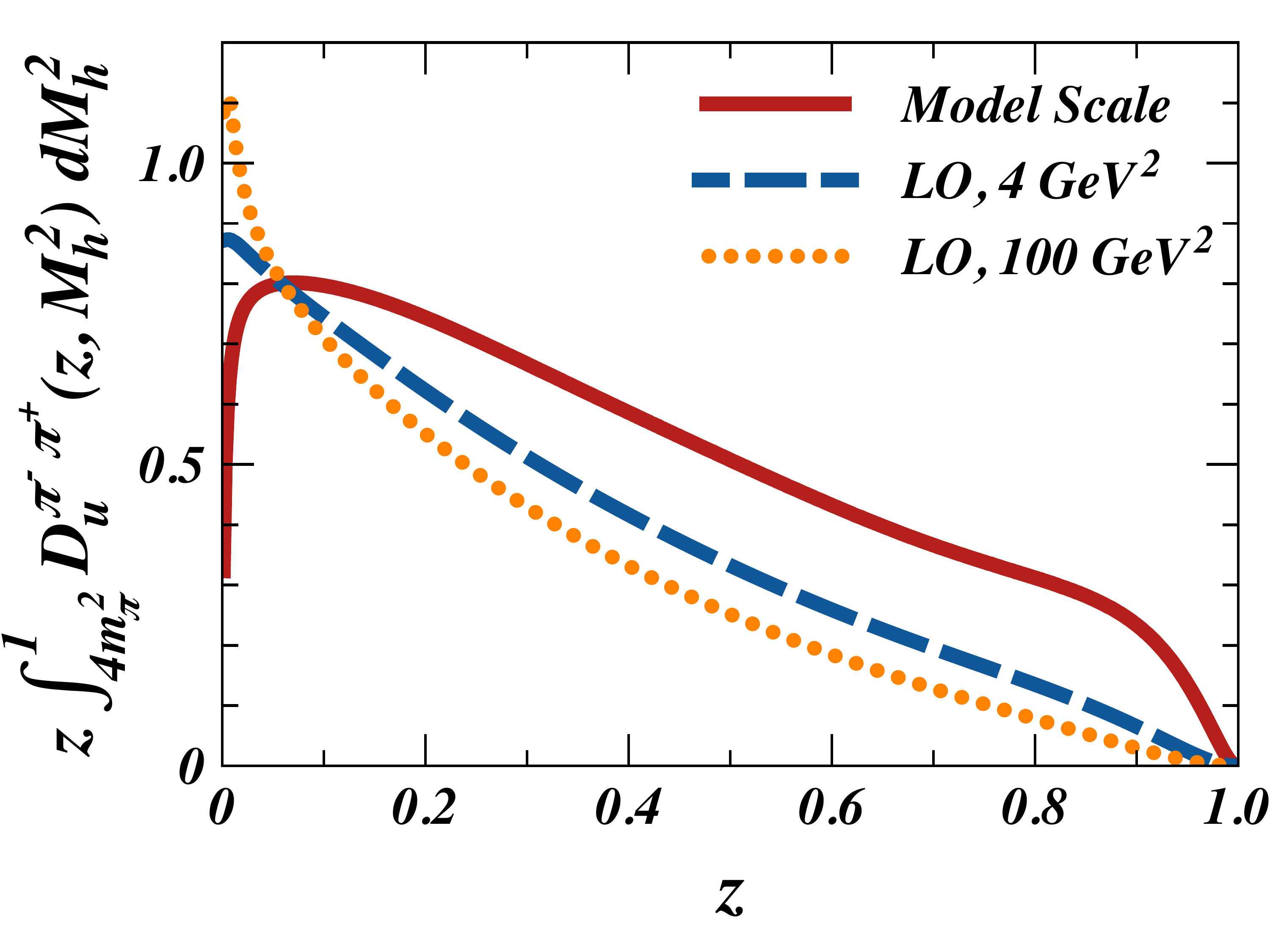}
}
\caption{The comparison of the results for $D_u^{\pi^+\pi^-}$ at model scale (red solid line) with those evolved  to $4~\Gs$ (blue dashed line) and $100~\Gs$ (orange dotted line) when (a) integrated over $z$ and (b) integrated over $M_h^2$.}
\label{PLOT_DFF_EVOL}
\end{figure}

 This evolution equation at a fixed invariant mass squared of the hadron pair is formally exactly the same as that for the single hadron fragmentation function. Thus we employ the evolution software QCDNUM~\cite{Botje:2010ay} to evolve our model calculated DFFs at fixed values of $M_h^2$ to the final scale of interest. The results for the evolution of $D_u^{\pi^+\pi^-}$ are shown in Fig.~\ref{PLOT_DFF_EVOL}. Here the model scale is taken to be the usual NJL-jet scale of $0.2~\Gs$, and the results are evolved to $4~\Gs$ and $100~\Gs$. The plots in Fig.~\ref{PLOT_DFF_EVOL}(a) show that QCD evolution of the DFF, integrated over an interval of $z$ between $0.2$ and $0.8$, leaves the overall shape of the $M_h^2$ dependence unchanged, reducing the overall magnitude and the relevant strength of different VM peaks. Such behaviour is easy to understand, as under evolution the strength of the DFF functions shifts to smaller values of $z$, as can be clearly seen from plots in Fig.~\ref{PLOT_DFF_EVOL}(b) depicting the same results integrated over a range of values for $M_h^2$.

\section{Conclusions and Outlook}
\label{SEC_CONC}

 In this work we have presented our NJL-jet model calculations of unpolarised  dihadron fragmentation functions of both light and strange quark to pairs of pions and kaons. For this the NJL-jet model and the corresponding Monte Carlo framework have been extended to extract these DFFs using their probabilistic interpretation. The important contributions from the production of several vector mesons and their strong decays were included, as they give a very large contribution to the DFF channels that correspond to their decay products - see also Ref.~\cite{Matevosyan:2013nla}. Thus, the two- and three-body decays of these resonances were included carefully in the model, using the experimental and phenomenological developments about these decays gained in precise experiments at $e^+e^-$ colliders~\cite{Achasov:2001hb}.  The results of the extensive MC simulations allowed us to extract the DFFs with insignificant numerical errors, where our studies showed that $8$ hadron emission steps in each quark hadronization chain are needed to achieve satisfactory convergence in the region $z\gtrsim 0.2$.
 
 The results for $u\to\pi^-\pi^+$ have confirmed our earlier findings of Ref.~\cite{Matevosyan:2013nla}, that the strong decays of the vector mesons play a crucial role in the DFFs of pseudoscalar mesons, both by increasing the magnitude and by shaping the invariant mass spectrum of the pair. This has also been confirmed for the DFFs of at least one final kaon produced by decays of $K^*$ and $\phi$ vector mesons as well. This behaviour can be explained using the large combinatorial factors involved in  counting the pairs produced by the vector meson decays. Moreover, the enhancement of the DFF magnitude is apparent even in the regions far from the invariant mass peaks of VMs and in the channels where production from decays alone is not possible (such as the similarly charged pairs). This is a general feature of DFFs, as we have shown in Ref.~\cite{Matevosyan:2013nla} by also using PYTHIA generated events for calculating DFFs.
 
  These calculations provide a wealth of information about DFFs using an effective quark model. These can be useful in empirical extractions of DFFs from data by providing indications on the validity of various assumption on their behaviour and symmetries. Moreover, experimental measurements of invariant mass profile of two hadron semi-inclusive production will be crucial in determining the relevant resonances that need to be included in the model description of DFFs. These measurements of both extended and "collinear" DFFs can as well serve as extensive tests of various hadronization models, including the NJL-jet model.
  
   The future development of our model will allow us to extract the interference dihadron fragmentation functions (IFF) by considering the fragmentation of a transversely polarised quark. The knowledge of these IFFs is crucial in extraction of transversity PDF from semi-inclusive two hadron production experiments. Such model calculations of IFFs and comparison with existing parametrizations of experimental data (Ref.~\cite{Courtoy:2012ry}) will help to understand the relevant aspects of hadronization for producing these chiral-odd functions. Another important task will be to build a complete software package for evolution of both extended and "collinear" DFFs using the evolution codes used in this work and in Ref.~\cite{Casey:2012hg}. Using the software and the sets of our (or any other) model calculated DFFs would allow users to easily compare the results with those extracted from experiment, providing critical tests for our assumptions and approximations used in the model building.   
    
\section*{Acknowledgements}

This work was supported by the Australian Research Council through Grants No. FL0992247 (AWT), No. CE110001004 (CoEPP), and by the University of Adelaide. W.B. acknowledges support by the Grant in Aid for Scientific Research (Kakenhi) of the Japanese Ministry of Education, Culture, Sports, Science and Technology, Project. No. 20168769. 


\bibliography{fragment}
\bibliographystyle{apsrev}

\end{document}